\documentclass{aa}  
\usepackage{graphicx}
\usepackage{longtable}
\usepackage{lscape}
\usepackage{txfonts}
\begin{document}
\title{The chemical composition of planetary nebulae and HII regions in \\ \object{NGC\,3109} \thanks{Based on observations collected at the European Southern Observatory, VLT, Paranal, Chile, program ID 076.B-0166. }~\thanks{This paper includes data obtained with the 6.5-m Magellan Telescopes located at Las Campanas Observatory, Chile.}}
\author{Miriam Pe\~na\inst{1}~\thanks{Part of this work was done while at DAS, Universidad de Chile, during a sabbatical year.}
\and
Gra\.zyna Stasi\'nska\inst{2}
\and
Michael G. Richer\inst{3}
}
\offprints{M. Pe\~na}
\institute{Instituto de Astronom{\'\i}a, Universidad Nacional Aut\'onoma de M\'exico, Apdo. Postal 70264, M\'ex. D.F., 04510 M\'exico\\
\email{miriam@astroscu.unam.mx}
\and
LUTH, Observatoire de Paris, CNRS, Universit\'e Paris Diderot; Place Jules Janssen 92190 Meudon, France\\
\email{grazyna.stasinska@obspm.fr}
\and
Instituto de Astronom{\'\i}a, Universidad Nacional Aut\'onoma de M\'exico, Apdo. Postal 877,  Ensenada, B.C., M\'exico\\
\email{richer@astrosen.unam.mx}
}
\date{Received 13/09/2007; accepted 28/09/2007}


\abstract
{}
{We  present deep  spectrophotometry for a sample of 8  planetary nebulae (PNe) and 12 HII regions distributed throughout the dwarf irregular galaxy NGC\,3109, in order to analyze the chemical composition of both types of nebulae. }
{We describe the observations and data reduction, and present line intensities for the nebular emission lines detected. The physical conditions and the abundances of He, O, Ne, N, S and Ar are derived, using the classical T$_{\rm e}$-based method. 
We confirm our previous identification of PNe and HII regions based on photometry, except for one object, which we argue is a compact HII region rather than a planetary nebula. }
{We find that the chemical composition of the interstellar medium in NGC\,3109, as sampled by its HII regions, is remarkably uniform. The oxygen abundance is log O/H + 12 =   7.77 $\pm$ 0.07 in this galaxy, as compared to 8.05 $\pm$ 0.09 for the Small Magellanic Cloud  (for which we rederived the metallicity in a homogeneous way).
PNe show significantly higher oxygen abundances in NGC\,3109: log O/H + 12 = 8.16 $\pm$ 0.19. Similarly to what has been suggested for some of the PNe in the Magellanic Clouds and other metal-poor galaxies, we argue that oxygen in the PNe  in  NGC\,3109 is affected by dredge up in their  progenitors. This could also be  the case for neon,  although the uncertainties for this  element are bigger.}  
{From our analysis, we conclude that oxygen and neon are not always a safe indicator of the chemical composition of the interstellar medium at low metallicities. 
An alternative to the O and Ne enrichment in PNe is that the low metallicity in HII regions has been caused by dilution of the interstellar medium due to an interaction with a neighboring galaxy about a Gyr ago. 
 The excitation patterns of the PNe in NGC\,3109 are very different from the excitation patterns of PNe in other galaxies.   This issue needs to be investigated further, as it implies that the evolution of  PNe depends upon the properties of their progenitor stellar populations, which vary from galaxy to galaxy.  This should affect the planetary nebula luminosity function and its use as a distance indicator.

Regarding individual objects, we find that the planetary nebula  named PN\,14 shows clear Wolf-Rayet features,  very low excitation and high density. Thus, it is similar to some of the galactic PNe ionized by  late [WC] stars.}

\maketitle
\section{Introduction}

Much effort has been made over the years to analyze the properties of the dwarf galaxies in the Local Group and its vicinity, as these objects hold important clues for stellar population studies. Dwarf irregulars, in particular, allow detailed studies of their different stellar populations, and through analysis of their stars, planetary nebulae (PNe) and HII regions, one can address questions such as how elemental abundances have evolved with time or what is the large-scale distribution of elements in the interstellar medium (ISM) of the galaxy.  

Recently, Leisy \& Dennefeld (2006) published a comprehensive study of abundances in PNe of the Magellanic Clouds.  They argued that, in those systems,  neither oxygen nor neon observed in the PNe can be used to derive the initial composition of the progenitors stars, since they are affected by nuclear processing during their evolution. On the other hand, Richer \& McCall (2007) analyzed the chemical compositions of \emph{bright PNe} and HII regions in dwarf irregular galaxies, and concluded that neither oxygen nor neon abundances in bright PNe are likely to be affected by the nucleosynthesis during the evolution of their progenitor stars.  In two objects, one in Sextans A and another in NGC 6822, they note that oxygen seems to have been dredged up, but they consider those cases as exceptions.  Obviously, a larger sample of good quality data, especially at low metallicity, would help to understand  the apparent contradiction between these studies.

In a previous paper (Pe\~na et al. 2007; hereafter Paper I), we published a catalog of compact HII regions and planetary nebula (PN) candidates in the Local Group  galaxy NGC\,3109 (DDO 236). This is the most complete catalog available for [\ion{O}{iii}] 5007 emitting objects, presenting accurate coordinates for 20 PN candidates and 40 HII regions.  This catalog also includes nebular magnitudes in [\ion{O}{iii}]5007 (the [\ion{O}{iii}]5007 fluxes reported here were used to calibrate these nebular magnitudes) and the magnitudes of the central stars.  Here,  we present a spectroscopic analysis of a substantial subsample of those objects.  We aim to analyze the large-scale spatial distribution of the heavy elements in the ISM of NGC\,3109 and to compare the chemical composition of the ISM at the present moment with its composition at the epoch represented by the PN population, presumably up to a few Gyrs ago.  We also analyze the relative abundances of elements, in particular the Ne/O ratio, to try to understand if these elements are truly representative of the ISM abundances at the time the PNe formed. A comparison with similar data for the Small Magellanic Cloud (SMC) is also discussed.   

NGC\,3109, an irregular late-type spiral (classified as SB(s)m  by de Vaucouleurs et al. 1991), is the dominant member of the Antlia-Sextans  group, located  just at the edge of the Local Group.   The galaxy has a cigar-like appearance  aligned almost E-W, with an extension of 17.4 $\times$3.5 arcmin.   It shows some traces of spiral structure (Demers et al. 1985). From recent studies of the Cepheid population, Soszy\'nski  et al. (2006) derived a distance modulus of 25.571$\pm$0.024 (1.30$\pm$0.02 Mpc) in coincidence with previous determinations. 

NGC\,3109 presents significant star formation activity, showing numerous extended and compact HII regions, as well as large scale ionized shells, super-shells and low brightness filaments.   Catalogs of its HII regions and extended ionized nebulae have been published by Hodge (1969),  Bresolin et al. (1993), and Hunter et al. (1993), among others.

The  spectrophotometric data presented here were obtained mostly with the Focal Reducer Spectrograph \#1 (FORS1) attached to the Very Large telescope (VLT) of the European Southern Observatory (ESO). The spectrograph was used in  multi-object mode (MOS), allowing us to obtain data for  12 HII regions and 7 PNe simultaneously, distributed across the galaxy. An additional PN was observed with the 6.5-m Magellan Telescope (Clay) and the high resolution spectrograph MIKE at Las Campanas Observatory. 

In Sect. 2, we present the  observations and data reduction procedures.   In Sect. 3, we present the reddening corrected line intensities for PNe and HII regions in NGC\,3109 and we comment on some differences in the spectral characteristics of the two groups of objects. In Sect. 4, we derive the physical conditions and chemical abundances of the nebulae and discuss the reliability of the results.  For comparison purposes, we also derive chemical abundances in the same way for PNe and HII regions in the SMC. In Sect. 5, we discuss the abundance patterns of the PNe and HII regions in NGC\,3109.  In Sect 6, we compare the nebulae in NGC\,3109 with those in the SMC.  In Sect. 7, we summarize our main conclusions. In the appendix, we discuss some interesting individual objects from our sample of PNe. 

\section{Observations and data reduction}

FORS1 in MOS mode was employed on January 28,  2006. As usual, the atmospheric dispersion was compensated by the instrument's linear atmospheric dispersion compensator. The grisms 600B+12 and 600V+94 were employed and the resulting maximum wavelength coverage was of 3450-5900 \AA\ at a dispersion of 1.2 \AA/pix in the blue and of 4650-7100 \AA\ at a dispersion of 1.18 \AA/pix in the red.  However, as a consequence of the different spatial positions of  MOS slitlets within the field of view, the exact wavelength coverage varied from object to object.  The minimum wavelength range spanned 3800 to 6600 \AA, but extended over 3700 to 6800 \AA \ for most objects.   

The large field of FORS1 ($6.8\times 6.8$ arcmin) allowed us to observe the whole galaxy in only two frames. 
In each field, the 19 slitlets of FORS1-MOS were accommodated. MOS slits  have a length of 20$''$-24$''$ and an adaptable width.  For all of our objects, we used a slit width of 1.7$''$ in order to include all the emission from compact sources and as much as possible for the more extended nebulae.  The seeing during the night  varied between 0.9$''$ and 1.5$''$.  For each mask and spectral range we obtained two or three frames in order to permit the removal of cosmic rays. The total exposure times were long enough to obtain faint emission lines important for plasma diagnostics. Table 1 presents a summary of the observations.

Our targets for spectroscopy were selected among the [\ion{O}{iii}] 5007-emitting objects found in the pre-imaging analysis.  (Pre-imaging was obtained in service mode on November 29 and December 1, 2005, program ID 076.B-0166(A), and constituted the basis for Paper I.)  The objects observed spectroscopically  were  the brightest PN candidates, namely PN\,3, PN\,4, PN\,7, PN\,11, PN\,13, PN\,14 and PN\,17, and the compact HII regions \# 4, 7, 11, 15, 17, 20, 30, 31, 32, 34, 37 and 40. These names  correspond to the ones given in Tables 2 and 3 of Paper I. 

\begin{table}
\caption{Summary of spectroscopic observations$^{a}$.}
\begin{tabular}{cccrr}
\hline \hline 
field & grism & filter& exp. time \\
\hline 
W    & 600B+12 & none& 3$\times$1800 s\\
W   & 600V+94 & GG375+30 & 2$\times$1200\,s, 1900\,s \\
E   & 600B+12 & none & 2$\times$1800\,s, 304\,s \\
E & 600V+94 & GG375+30 & 1800\,s, 900\,s \\
\hline
\multicolumn{4}{l}{$a$ Observing date: January 28, 2006}
\end{tabular}
\end{table}

\begin{figure*}
\includegraphics[width=9cm]{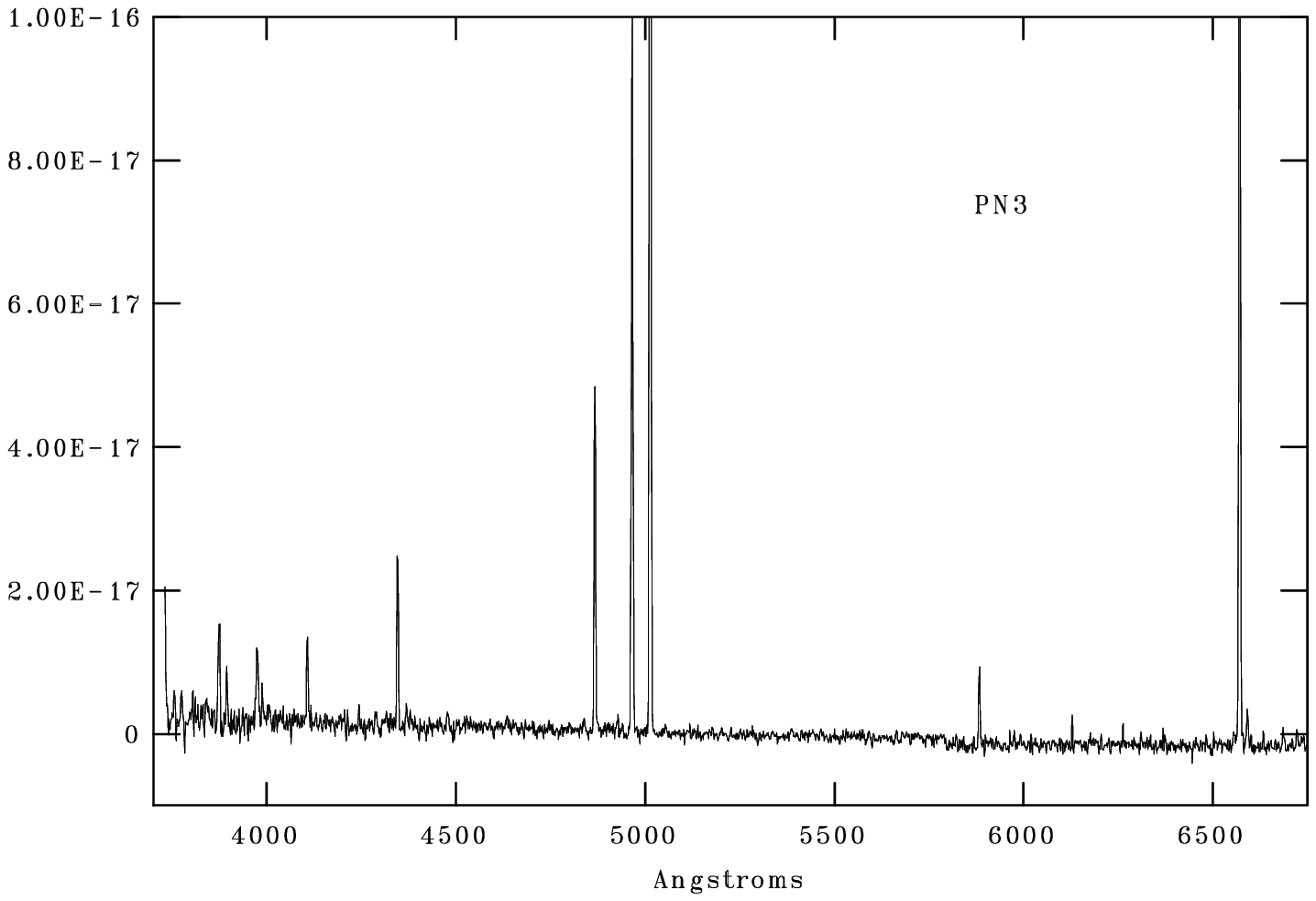} 
\includegraphics[width=9cm]{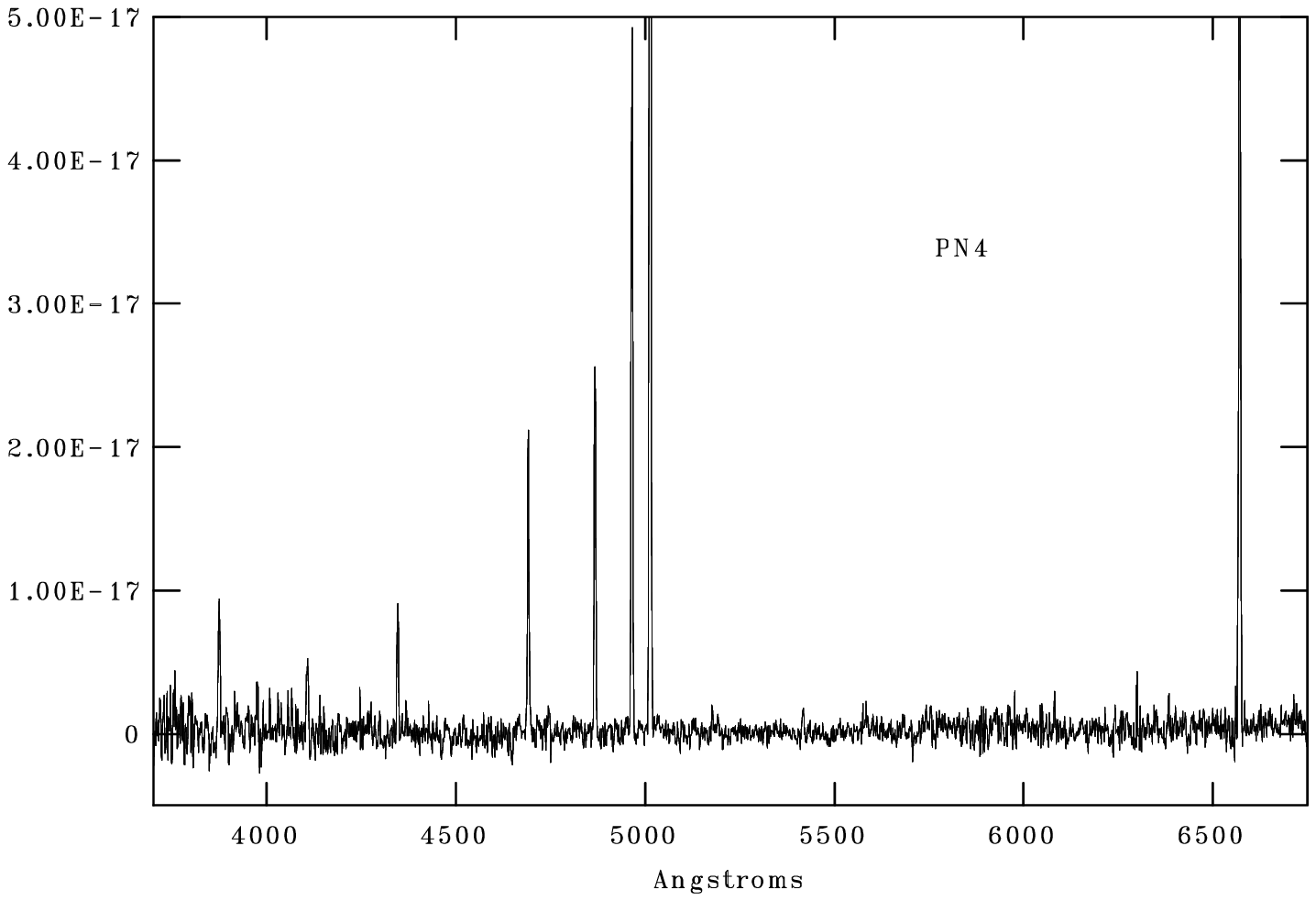}
\includegraphics[width=9cm]{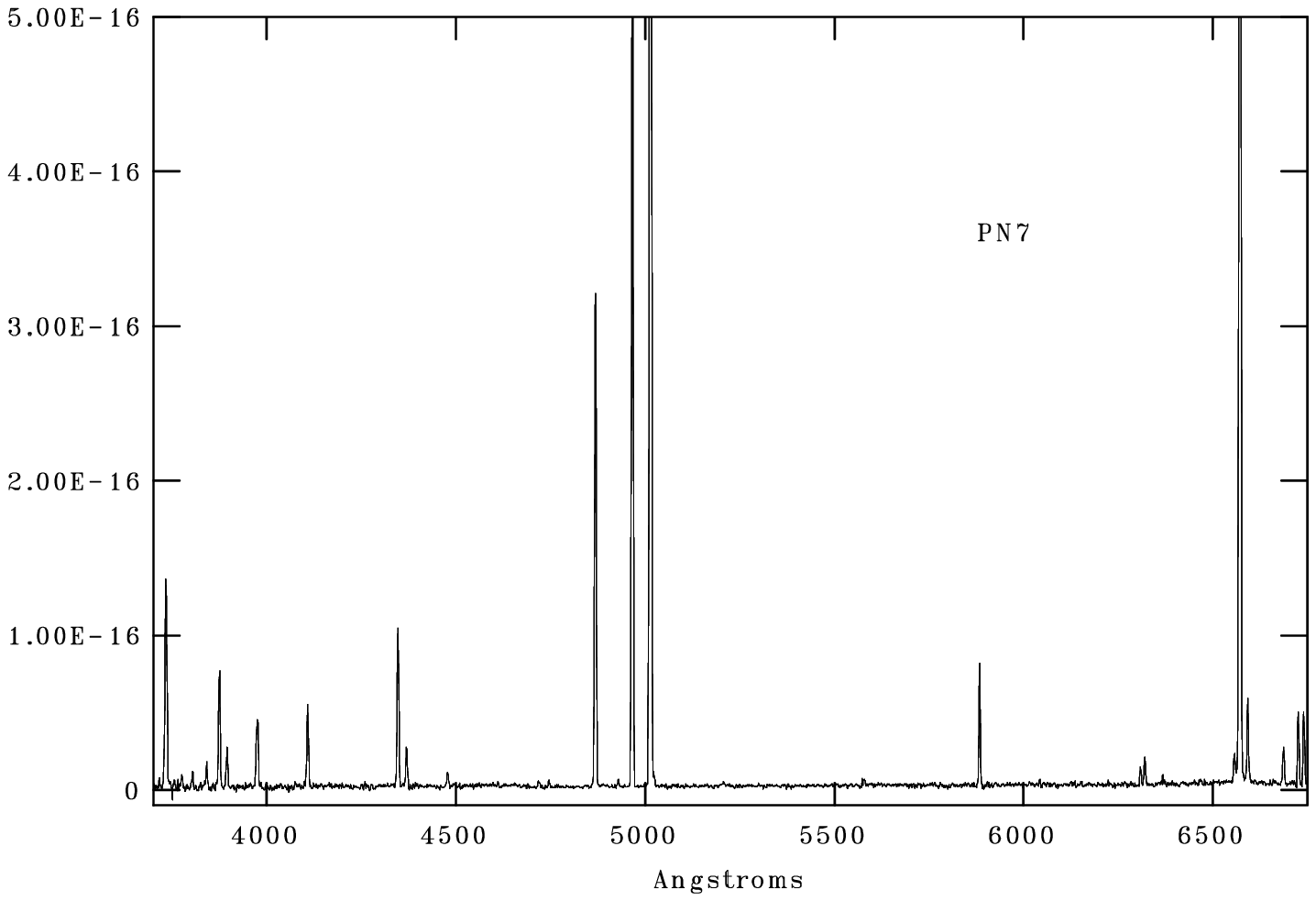}
\includegraphics[width=9cm]{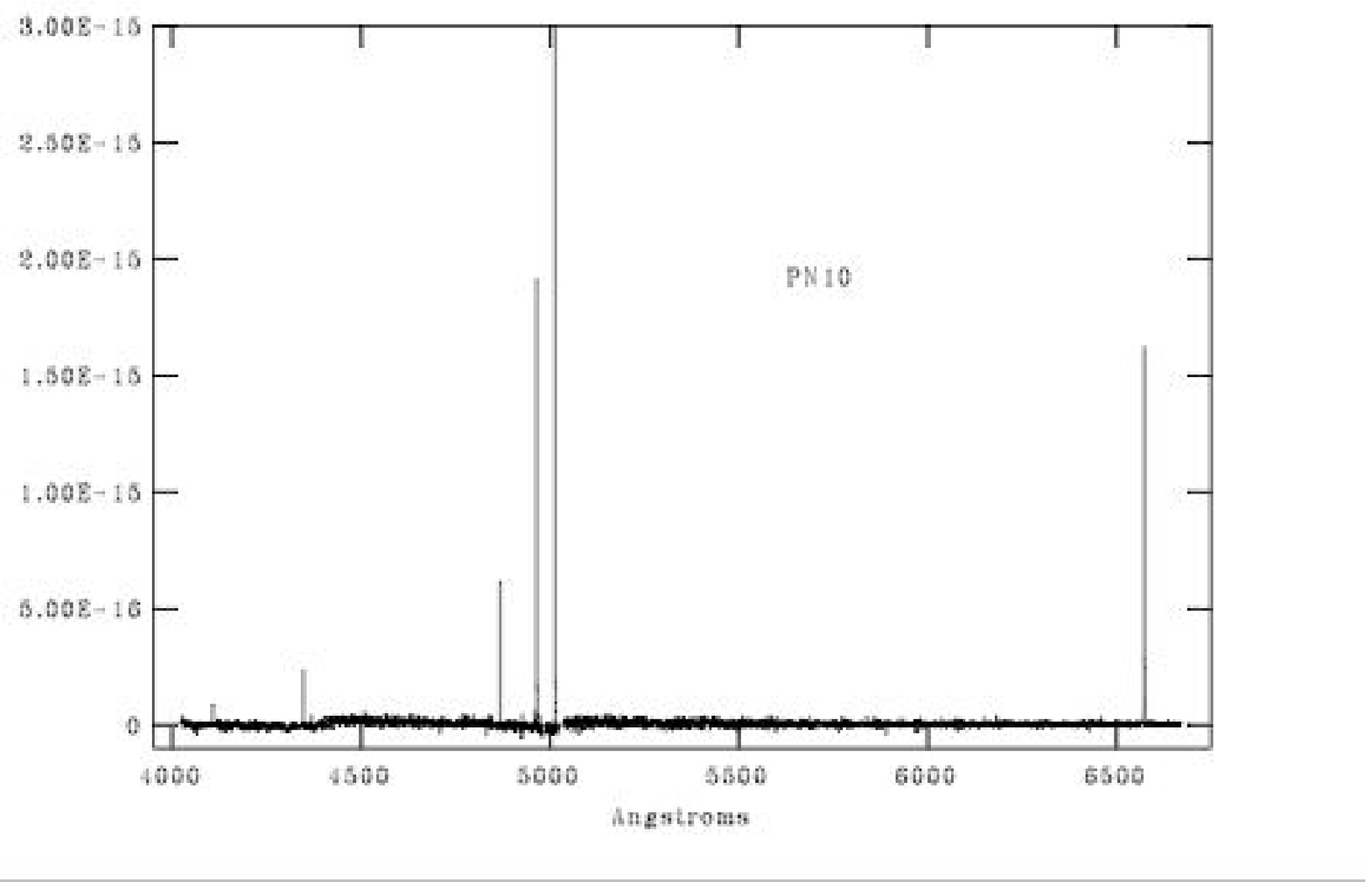}
\includegraphics[width=9cm]{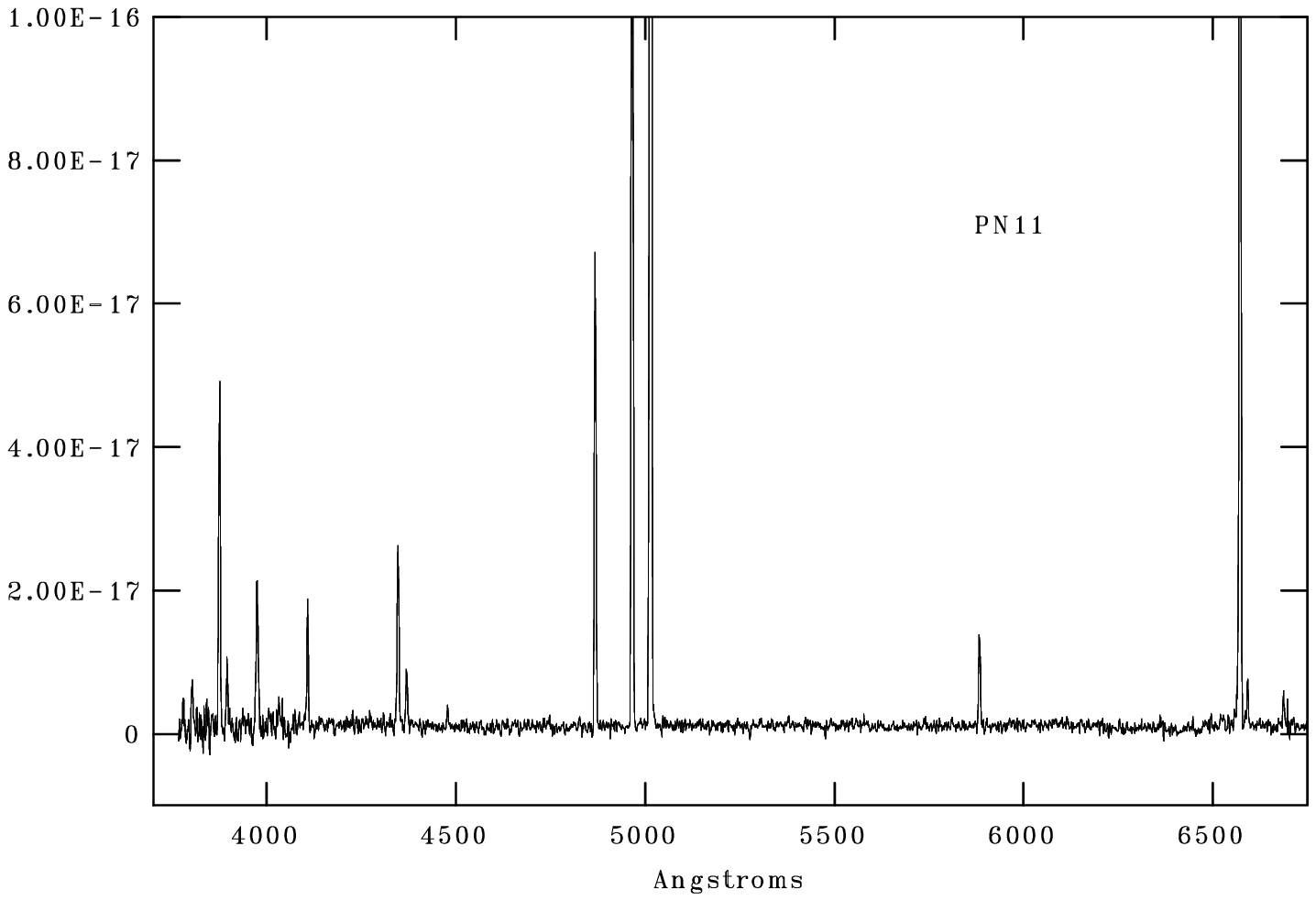}
\includegraphics[width=9cm]{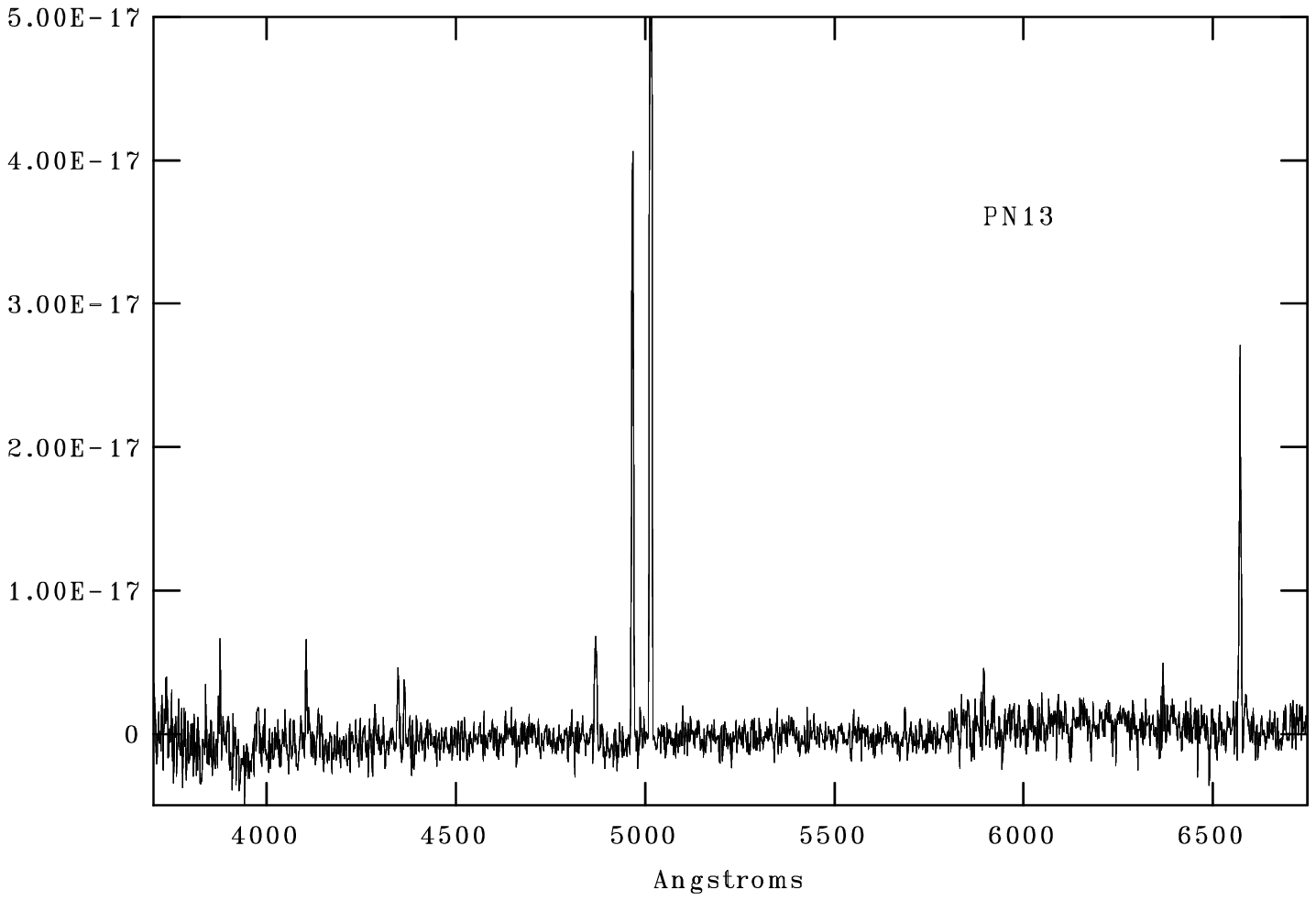}
\includegraphics[width=9cm]{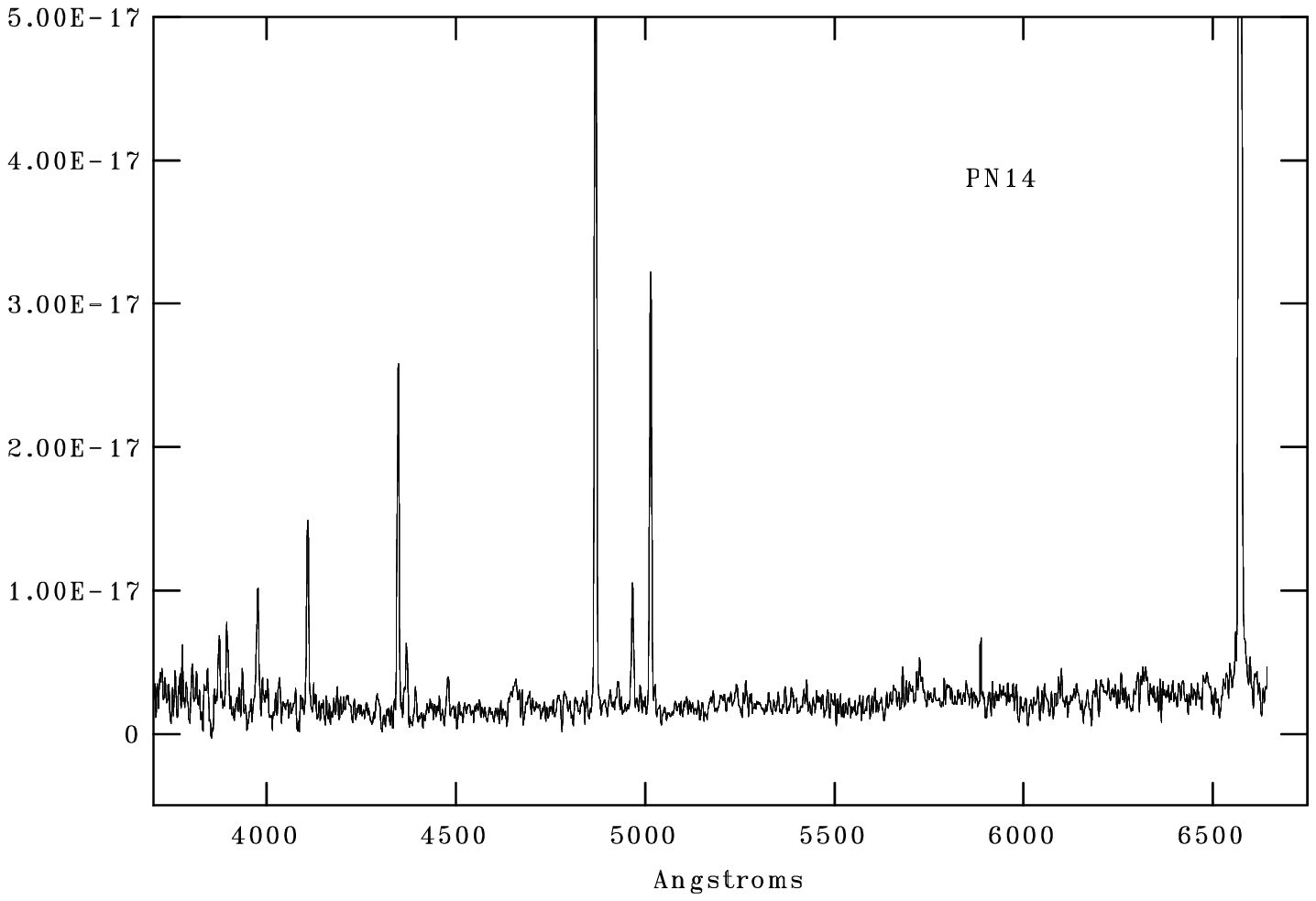}
\includegraphics[width=9cm]{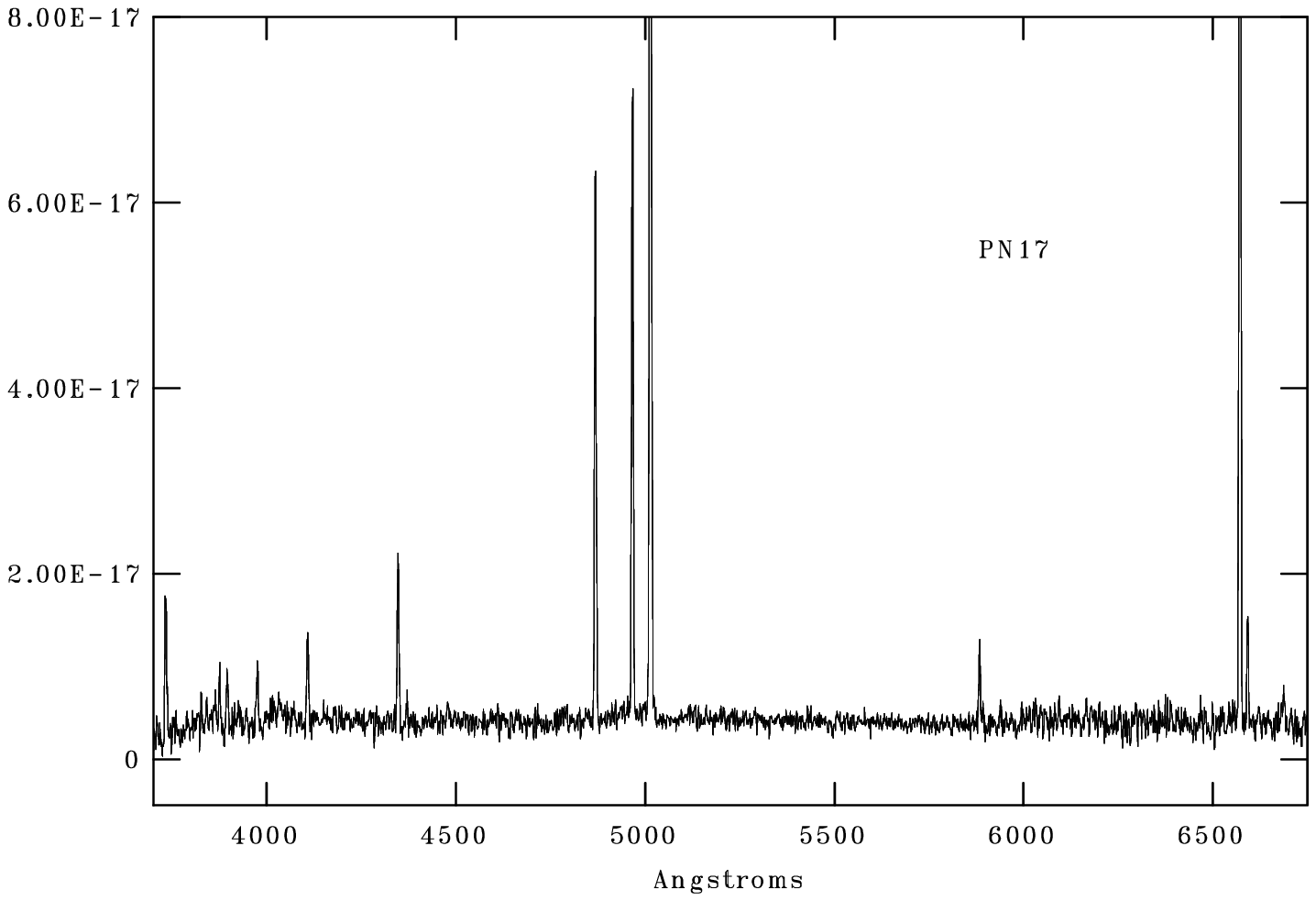}
\caption{Calibrated spectra obtained for the PNe. Several echelle orders are shown for PN\,10 (observed with the Magellan Telescope and MIKE spectrograph).\label{fig1}}
\end{figure*}

\begin{figure*}
\includegraphics[width=9cm]{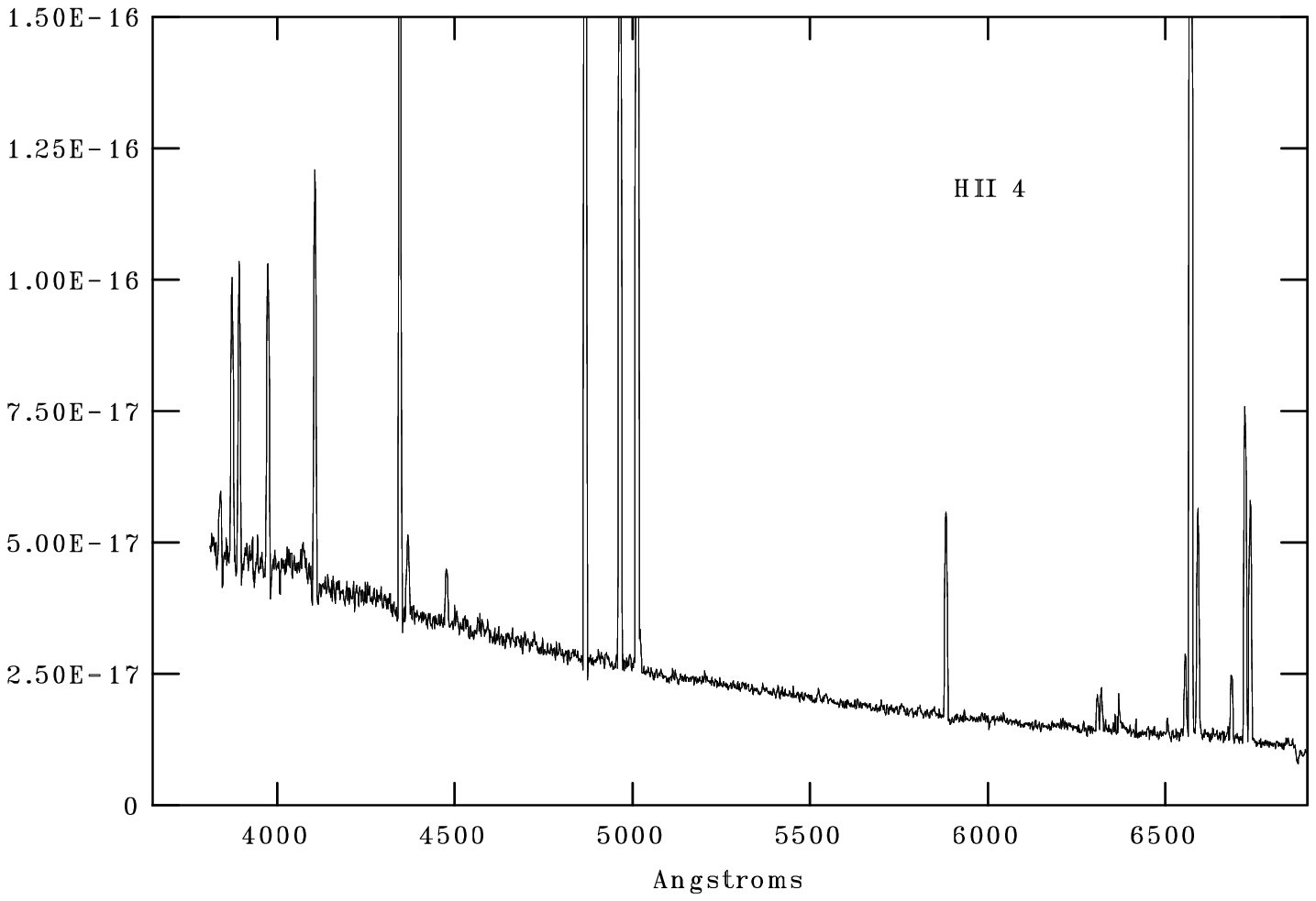} 
\includegraphics[width=9cm]{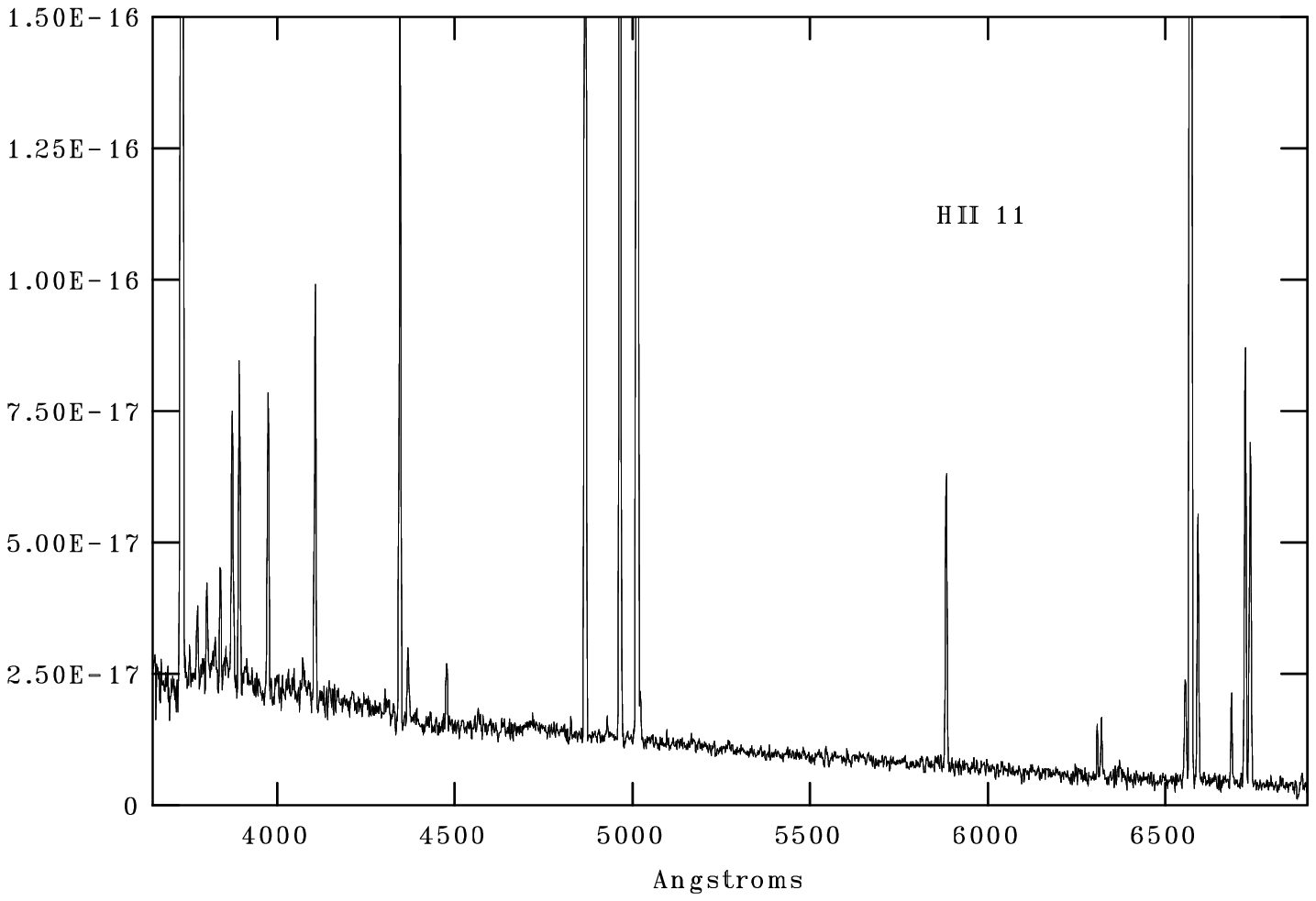}
\includegraphics[width=9cm]{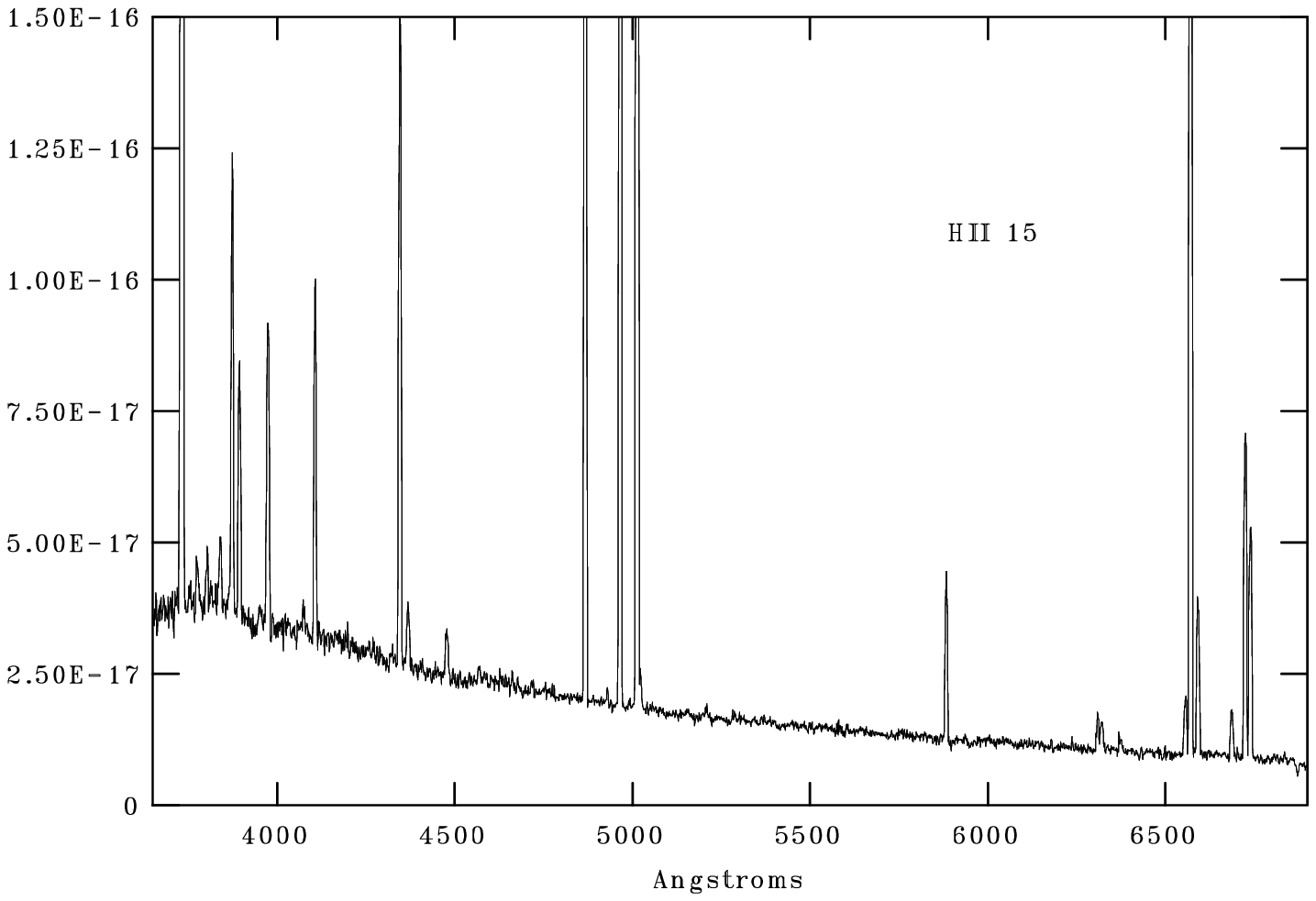}
\includegraphics[width=9cm]{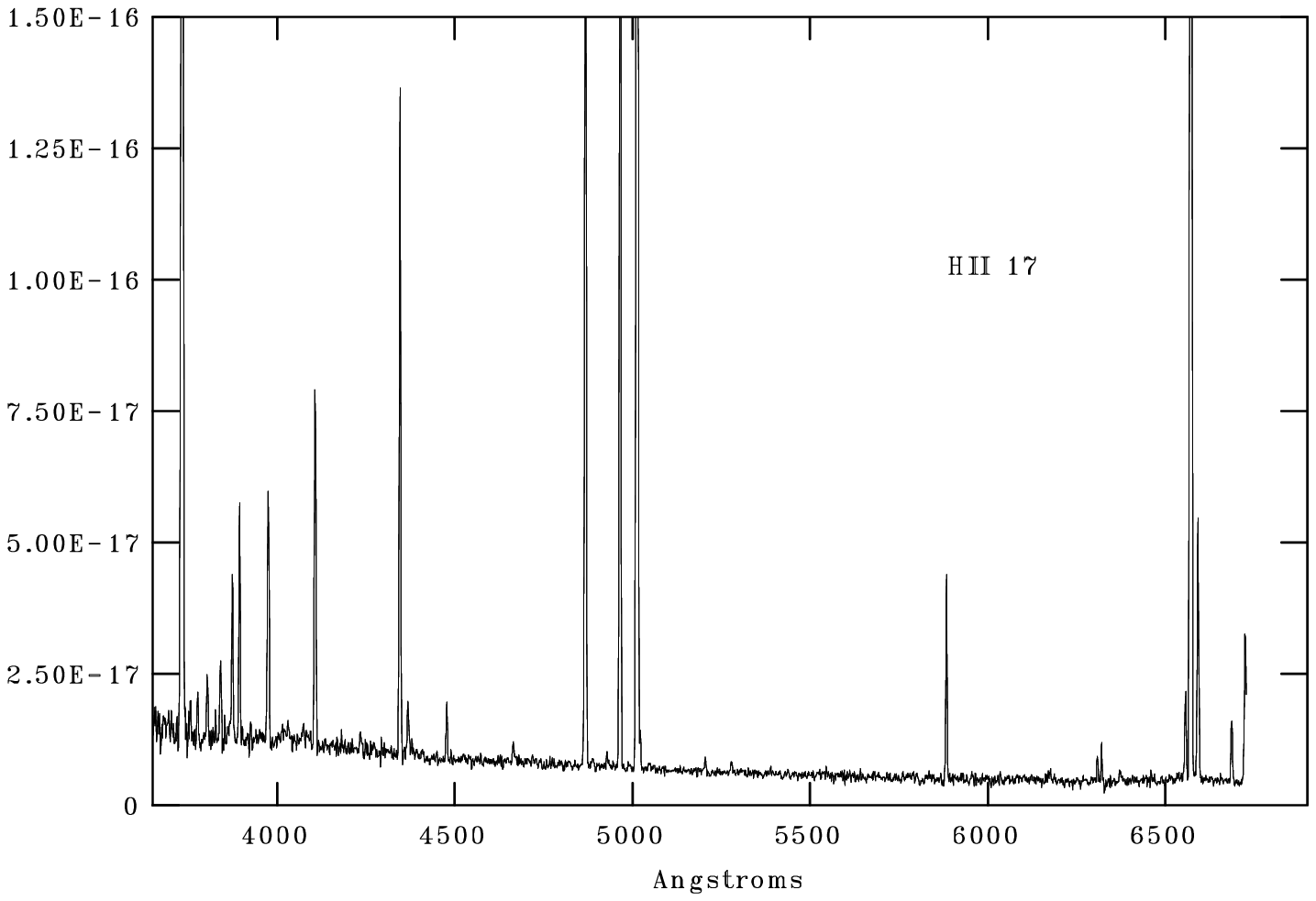}
\includegraphics[width=9cm]{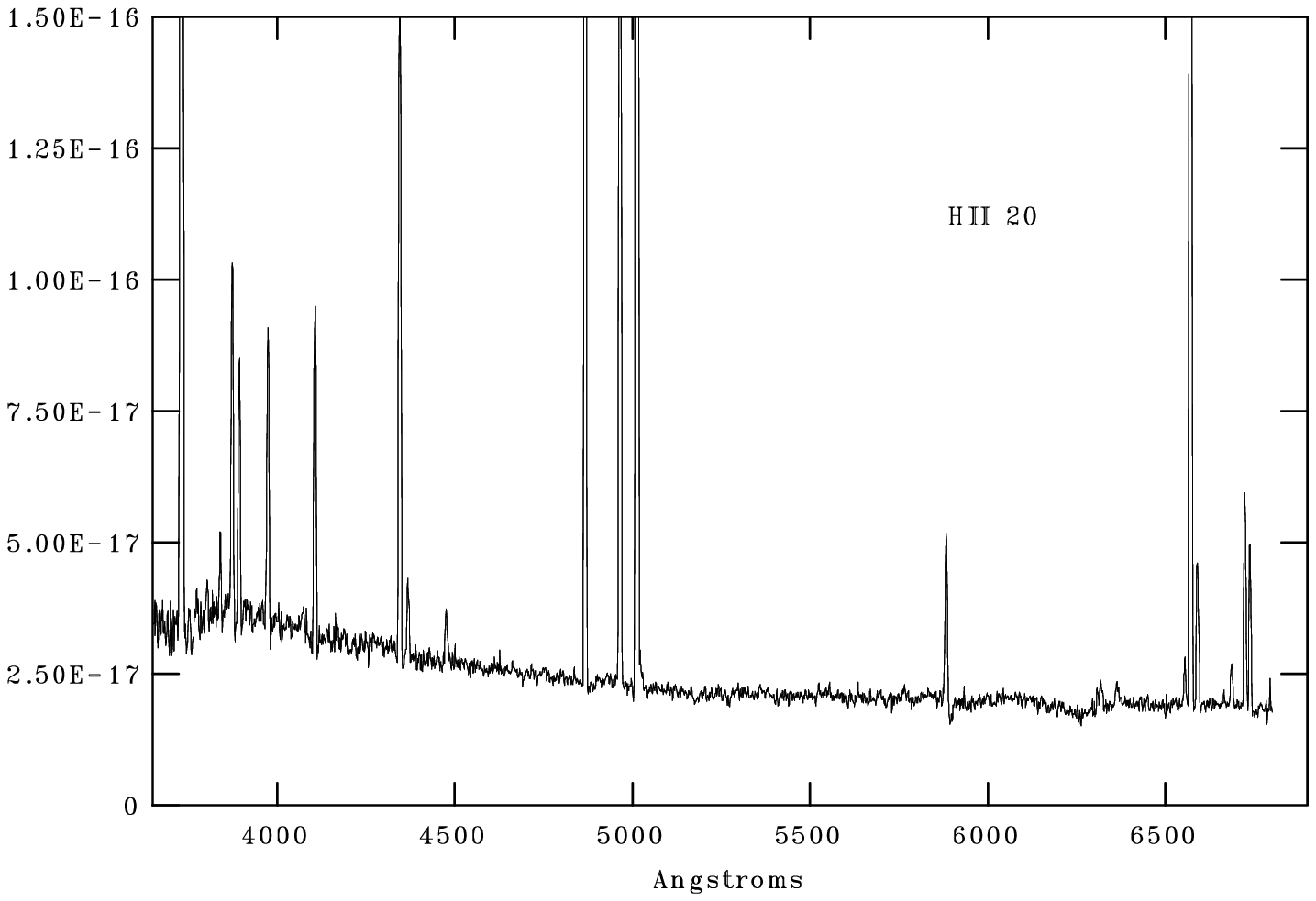}
\includegraphics[width=9cm]{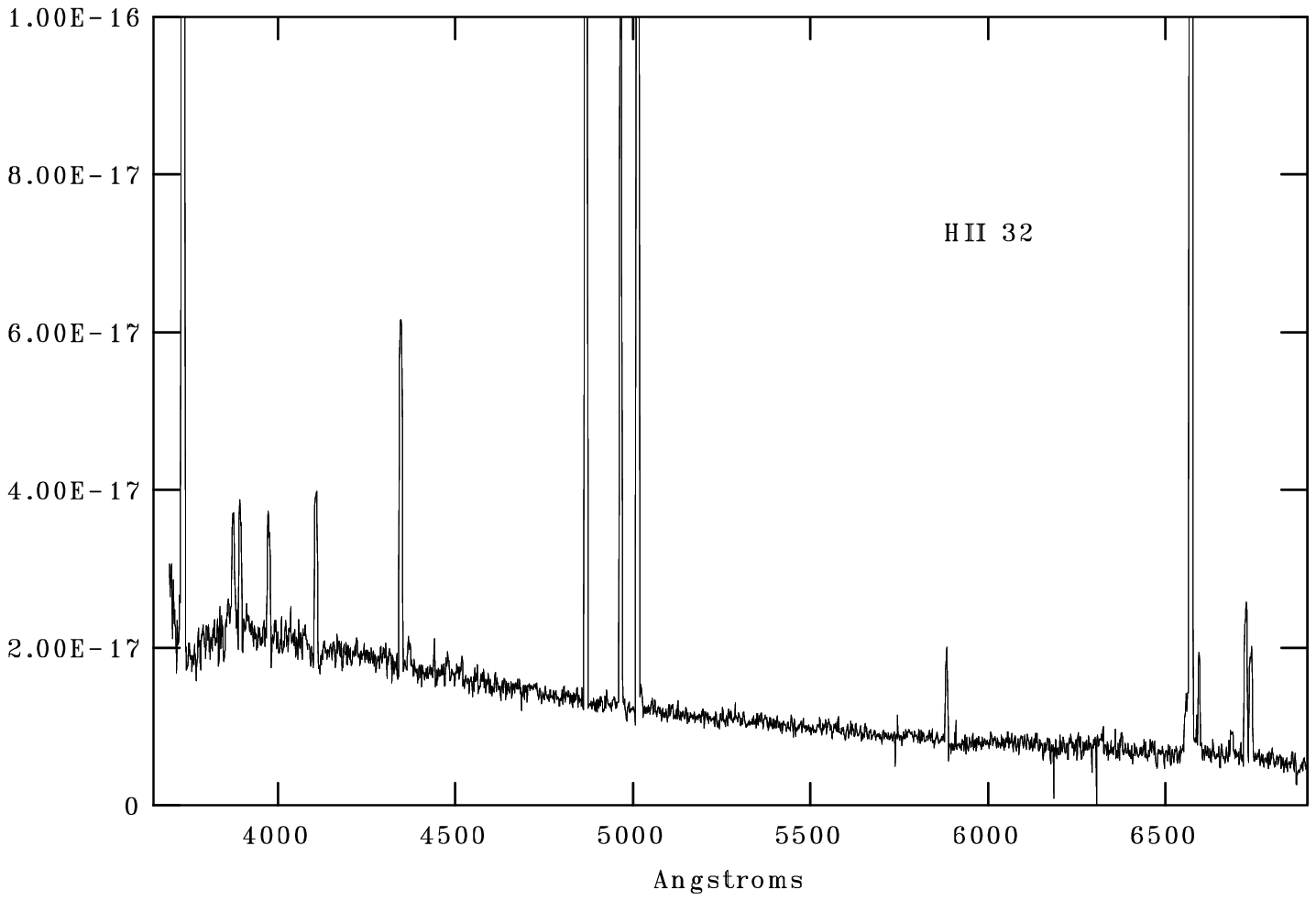}
\includegraphics[width=9cm]{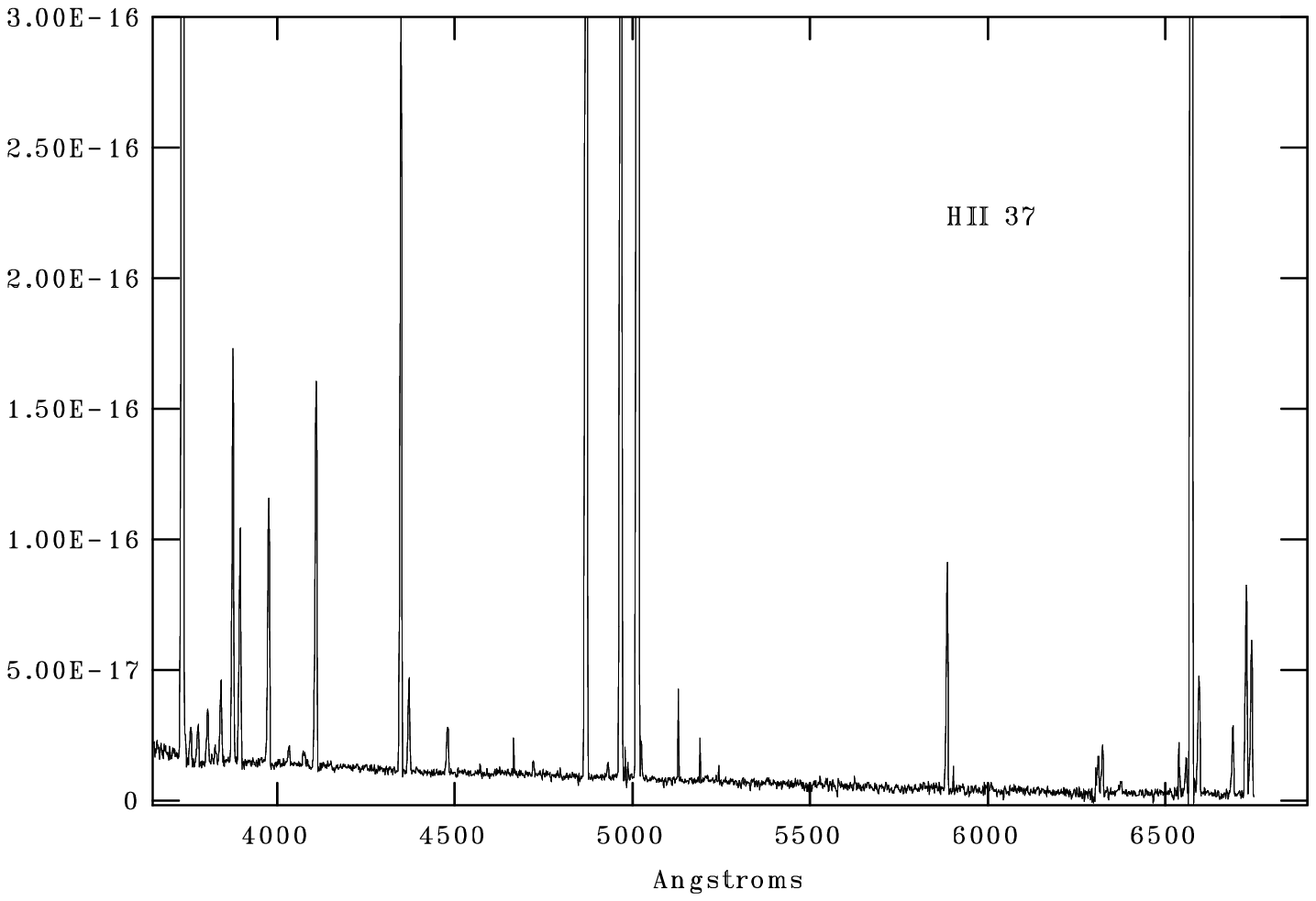}
\includegraphics[width=9cm]{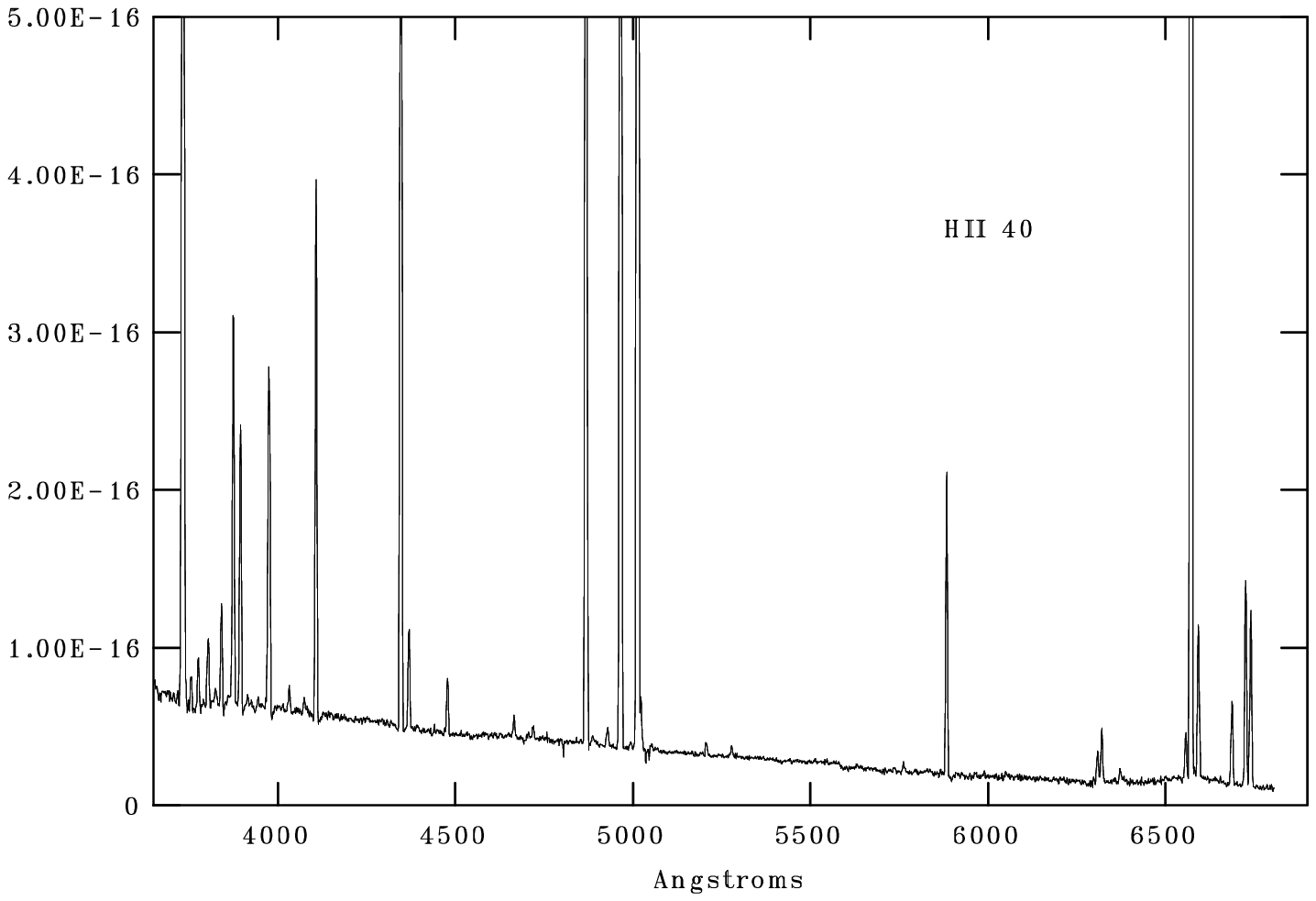}
\caption{Calibrated spectra obtained for some HII regions.  \label{fig2}}
\end{figure*}

\subsection {MOS data reduction}

Data reduction was performed using IRAF\footnote{IRAF is distributed by the National Optical Astronomy Observatories, which is operated by the Association of Universities for Research in Astronomy, Inc., under contract to the National Science Foundation.}  routines.
Calibration frames (bias, dome flat-fields and arcs for wavelength calibration) were acquired during the day following the normal procedures at ESO-VLT.  The bias frames were combined to produce one `zero' frame.  For each MOS mask, the flats were also combined to produce one flat-field for each spectral range. For flux calibration, the standard star HD\,49798 was observed.

The raw `science' frames for each mask (wavelength range) were combined to produce a unique frame with better signal-to-noise.  These frames were bias-subtracted and flat-fielded.  Then, the individual spectra were extracted using an extraction window large enough to include all of the emission from each object;  a width of about 2 arcsec was used for compact objects and up to 3 arcsec was used for the more extended ones. The length of the slitlets was always long enough to allow for sky subtraction, even for the HII regions since the most extended of those observed has a diameter of about 3 arcsec.

The calibration arcs  were extracted following the same procedures as the corresponding objects and were used to wavelength calibrate the spectra.   The combination of grisms and the slit width produced a spectral resolution of 7.8 \AA, as measured with the arcs (FWHM of lines). 

Figure 1 and Fig. 2 display the calibrated spectra for all  the observed PNe and for some of the HII regions, respectively. 

\subsection{Observations with the 6.5-m Magellan Telescope}

PN\,10, one of the brightest PNe in NGC\,3109, was observed at Las Campanas Observatory with the Clay 6.5-m telescope and the high resolution spectrograph MIKE the night of May 9, 2006.  The blue and red arms of the spectrograph  covered wavelength ranges of 3500-5000 \AA\ and 4900-9600 \AA, respectively. The total exposure time was 6 $\times$ 900\,s (1.5 hr in total) and the slit size was 1$\times$5 arcsec.  A Th-Ar lamp was used for wavelength calibration. The spectral resolution of these spectra is better than 0.1 \AA. The individual frames were combined to improve the signal-to-noise and to eliminate cosmic rays. The flux standards HD\,49798 and HR\,5501 were observed for flux calibration. Data reduction was performed with IRAF routines, using standard procedures.

\begin{table*} 
\caption{ Dereddened line intensities  HII regions and PNe, relative to H$\beta$=1.00.}
\label{}\centering\begin{tabular}{lccrrrrrrrrrr}
\hline
ion	 & 	$\lambda$ &  f($\lambda$)	 & 	HII\,4	 & 	HII\,7	 & 	HII\,11	 & 	HII\,15	 & 	HII\,17	 & 	HII\,20	 & 	HII\,30	 & 	HII\,31	 & 	HII\,32	 & 	HII\,34	 \\
\hline
$[\ion{O}{ii}]$& 	3727	 & 	 +0.255	 & 	 oor	 & 	 1.58 	 & 	 2.29 	 & 	 1.63 	 & 	 1.89 	 & 	 1.73 	 & 	 2.00 	 & 	 1.09 	 & 	 1.72 	 & 	 3.03 	 \\
H11  	 & 	3770	 & 	 +0.250	 & 	oor	 & 	$<$0.02 	 & 	 0.05 	 & 	 0.03 	 & 	 0.03	 & 	$<$0.03  & 	 0.03 	 & 	$<$0.03  & 	$<$0.02  & 	$<$ 0.02  \\
H10  	 & 	3798	 & 	 +0.245	 & 	oor	 & 	 0.02: 	 & 	0.06 	 & 	 0.03 	 & 	 0.04 	 & 	 0.03 	 & 	 0.04 	 & 	 0.06 	 & 	 0.02: 	 & 	$<$0.02 	 \\
H9   	 & 	3835	 & 	 +0.236 	 & 	0.04 	 & 	 0.05 	 & 	 0.07 	 & 	 0.05 	 & 	 0.05 	 & 	 0.04 	 & 	 0.04 	 & 	 0.05 	 & 	$<$0.02  & 	$<$ 0.02 \\
$[\ion{Ne}{iii}]$  & 	3868	 & 	 +0.223  & 	0.12 	 & 	 0.17 	 & 	 0.12 	 & 	 0.20 	 & 	 0.09 	 & 	 0.14 	 & 	 0.07 	 & 	 0.15 	 & 	 0.13 	 &      $<$ 0.02 \\
H8  	 & 	3889	 & 	 +0.223 	 & 	 0.11 	 & 	 0.11 	 & 	 0.12 	 & 	 0.11 	 & 	 0.11  	 & 	 0.11 	 & 	 0.11 	 & 	 0.11 	 & 	 0.10 	 & 	 0.10:	 \\
H7+[\ion{Ne}{iii}]& 	3970 &   +0.204 	 & 	 0.19 	 & 	 0.18 	 & 	 0.20 	 & 	 0.22 	 & 	 0.20 	 & 	 0.20 	 & 	 0.18 	 & 	 0.17 	 & 	 0.10 	 & 	 0.19 	 \\
\ion{He}{i}  & 	4026	 & 	 +0.197 	 & 	$<$ 0.01  & 	 0.01: 	 & 	 0.01: 	 & 	 $<$0.01  & 	 0.01: 	 & 	 $<$0.02  & 	 $<$0.01 & 	 $<$0.01 & 	 0.01	 & 	 $<$ 0.02 \\
$[\ion{S}{ii}]$   & 	4069	&  +0.187 	 & 	$<$ 0.01  & 	 $<$0.01  & 	 0.01 	 & 	 0.02 	 & 	 0.01: 	 & 	 0.02 	 & 	$<$0.01  & 	$<$0.01  & 	 0.02:	 & 	 $<$0.02  \\
H$\delta$   & 	4102	 & 	 +0.182 	 & 	 0.25 	 & 	 0.23 	 & 	 0.24 	 & 	 0.25 	 & 	 0.28 	 & 	 0.23 	 & 	 0.27 	 & 	 0.26 	 & 	 0.26 	 & 	 0.18 	 \\
H$\gamma$  & 	4340	 & 	 +0.125 	 & 	 0.47 	 & 	 0.47 	 & 	 0.46 	 & 	 0.46 	 & 	 0.47 	 & 	 0.47 	 & 	 0.47 	 & 	 0.47 	 & 	 0.47 	 & 	 0.37 	 \\
$[\ion{O}{iii}]$ & 	4363 & 	 +0.121 	 & 	 0.06 	 & 	 0.06 	 & 	 0.05 	 & 	 0.06 	 & 	 0.04 	 & 	 0.06 	 & 	 0.02: 	 & 	 0.05::  & 	 0.03 	 & 	$<$0.02  \\
\ion{He}{i}	 & 	4471 & 	 +0.078 	 & 	 0.03 	 & 	 0.04 	 & 	 0.04 	 & 	 0.04 	 & 	 0.04 	 & 	 0.04 	 & 	 0.02: 	 & 	 0.02 	 & 	 0.02 	 & 	$<$ 0.02  \\
\ion{He}{ii}$^{(a)}$  & 	4686	 & 	 +0.070	 & 	   	 & 	    	 & 	  	 & 	   	 & 	   	 & 	    	 & 	  	 & 	    	 & 		 & 		 \\
$[\ion{Ar}{iv}]$\,+\,\ion{He}{i}& 4712 & +0.039  & 	$<$ 0.01  & 	 0.01 	 & 	 $<$0.01 & 	 $<$0.01  & 	 0.01 	 & 	 0.01 	 & 	$<$0.01  & 	 $<$0.01  & 	$<$0.01  & 	$<$ 0.02  \\
$[\ion{Ar}{iv}]$ & 	4738	 & 	 +0.031  & $<$ 0.01  & 0.01 & $<$0.01 & $<$0.01  &  $<$0.01 	 & 	$<$0.01  & 	$<$0.01  & 	$<$0.01  & 	$<$0.01  & 	$<$0.02 \\
H$\beta$	 & 	4861	 & 	 +0.000 	 & 	1.00	 & 	 1.00	 & 	1.00	 & 	 1.00	 & 	 1.00	 & 	 1.00	 & 	 1.00	 & 	 1.00	 & 	 1.00	 & 	 1.00	 \\
\ion{He}{i}	 & 	4922	 & 	 $-$0.014 	 & 	 0.01: 	 & 	 0.01 	 & 	 0.01 	 & 	 0.01 	 & 	 0.01 	 & 	$<$0.01  & 	$<$0.01  & 	 0.01 	 & 	 0.01 	 & $<$ 0.02  \\
$[\ion{O}{iii}$] & 	4959	 & 	 $-$0.023	 & 	 0.79 	 & 	1.05 	 & 	0.85 	 & 	 1.21 	 & 	 0.87 	 & 	 1.30 	 & 	 0.42 	 & 	 0.91 	 & 	 0.65 	 & 	 0.15 	 \\
$[\ion{O}{iii}]$ & 	5007	 & 	 $-$0.033	 & 	 2.33 	 & 	 3.02 	 & 	 2.47 	 & 	 3.44 	 & 	 2.64 	 & 	 3.81 	 & 	 1.47 	 & 	 2.85 	 & 	 1.98 	 & 	 0.65 	 \\
\ion{He}{i}	 & 	5017	 & 	 $-$0.036 	 & 	 0.01	 & 	 0.02 	 & 	 0.02 	 & 	 0.02 	 & 	 0.02 	 & 	0.01 	 & 	 $<$0.02 & 	$<$0.01 & 	0.01 	 & 	 $<$0.02 \\
$[\ion{N}{i}]$ & 	5199	 & 	 $-$0.083 	 &   $<$ 0.01  &  $<$0.01 	 & $<$0.01 	 & 	$<$0.01 & 	 0.01	 & 	$<$0.01 & 	$<$0.01 & 	 $<$0.01 & 	$<$0.02 & 	$<$ 0.02 \\
$[\ion{N}{ii}]$ & 	5755	 & 	 $-$0.195	 &   $<$ 0.01  & $<$0.01  & 	$<$0.01 	 & 	$<$0.01 & 	$<$0.01 & 	 0.01 	 & 	$<$0.01 & 	$<$0.01  & 	$<$0.01  & 	$<$ 0.02  \\
\ion{He}{i}	 & 	5876	 & 	 $-$0.280	 & 	 0.10 	 & 	 0.10 	 & 	 0.11 	 & 	 0.13 	 & 	 0.11 	 & 	 0.14 	 & 	 0.11 	 & 	 0.12 	 & 	 0.10 	 & 	 0.11 	 \\
$[\ion{O}{i}]$ & 	6300	 & 	 $-$0.284 	 & 	 0.02 	 & 	 0.01:  & 	 0.02 	 & 	 0.03 	 & 	 0.01 	 & 	 0.01 	 & 	$<$0.01 & 	$<$0.01 & 	$<$0.01 & 	$<$ 0.02 \\
$[\ion{S}{iii}]$ & 	6311	 & 	 $-$0.286 	 & 	 0.02 	 & 	 0.03 	 & 	 0.02 	 & 	 0.03 	 & 	 0.02 	 & 	 0.02 	 & 	$<$0.02 & 	$<$0.01 & 	 0.01 	 & 	$<$ 0.02 \\
$[\ion{N}{ii}]$ & 	6548	 & 	 $-$0.300 	 & 	 0.05 	 & 	 0.02 	 & 	 0.05 	 & 	 0.05 	 & 	 0.06 	 & 	 0.03 	 & 	 0.02 	 & 	 0.02:  & 	 0.06 	 & 	 0.04 	 \\
H$\alpha$ & 	6562	 & 	 $-$0.322	 & 	 2.86 	 & 	 2.86 	 & 	 2.86 	 & 	 2.86 	 & 	 2.86 	 & 	 2.81 	 & 	 2.82 	 & 	 2.86 	 & 	 2.86 	 & 	 2.86 	 \\
$[\ion{N}{ii}]$ & 	6583	 & 	 $-$0.325 	 & 	0.14 	 & 	 0.08 	 & 	 0.11 	 & 	 0.13 	 & 	 0.13 	 & 	 0.12 	 & 	 0.09 	 & 	 0.05 	 & 	 0.18 	 & 	 0.17 	 \\
\ion{He}{i}	 & 	6678	 & 	 $-$0.339 	 & 	 0.04 	 & 	 0.04 	 & 	 0.03 	 & 	 0.04 	 & 	 0.03 	 & 	 0.04 	 & 	 0.03 	 & 	 0.03 	 & 	 0.02 	 & 	 0.02: 	 \\
$[\ion{S}{ii}]$ & 	6717	 & 	 $-$0.343 	 & 	0.19 	 & 	 0.14 	 & 	 0.19 	 & 	 0.28 	 & 	 oor	 & 	 0.17 	 & 	 0.14 	 & 	 0.13 	 & 	 0.26 	 & 	 0.26 	 \\
$[\ion{S}{ii}]$ & 	6731	 & 	 $-$0.344 	 & 	0.14 	 & 	 0.09 	 & 	 0.15 	 & 	 0.20 	 & 		 & 	 0.12 	 & 	 0.11 	 & 	 0.10 	 & 	 0.18 	 & 	 0.13 	 \\
\ion{He}{i}	 & 	7066	 & 	 $-$0.383 	 & 	 0.02 	 & 	 0.02 	 & 	oor	 & 	oor	 & 		 & 	 oor	 & 	 0.02 	 & 	 oor	 & 	 0.03 	 & 	 oor	 \\
$[\ion{Ar}{iii}]$ & 	7135	 & 	 $-$0.390 	 & 	 0.06 	 & 	 0.08 	 & 		 & 		 & 		 & 		 & 	 0.05 	 & 		 & 	 0.08 	 & 	   	 \\
$[\ion{O}{ii}]$ & 	7320	 & 	 $-$0.410 	 & 	 0.03 	 & 	oor 	 & 		 & 		 & 		 & 		 & 	oor	 & 		 & 	oor	 & 		 \\
$[\ion{O}{ii}]$ & 	7330	 & 	 $-$0.411 	 & 	 0.02 	 & 		 & 		 & 		 & 		 & 		 & 		 & 		 & 		 & 		 \\
\hline
 c(H$\beta$)    & 		 & 		        & 	0.12	 & 	0.08	 & 	0.1	 & 	 0.06 	 & 	0.33	 & 	0.06	 & 	0.18	 & 	0.06	 & 	0.03	 & 	0.11	 \\
\multicolumn{3}{l}{F(H$\beta$) (10$^{-16}$ erg cm$^{-2}$ s$^{-1}$)}   & 26.6 	 & 	25.2 	 & 	17.8 	 & 	 26.0 	 & 	16.0 	 & 	23.0 	 & 	11.0 	 & 	8.30 	 & 	13.2 	 & 	9.20 	 \\
\multicolumn{3}{l}{L(H$\beta$)\,/\,L$_\odot$}                         & 140	 & 	130	 &	94	 & 	140	 &	85	 & 	120	 & 	58	 &	44	 & 	70	 & 	49	\\
\hline
\multicolumn{12}{l}{ $a$ The \ion{He}{ii} 4686 line is not detected in any of the HII regions. Then this line appears blank for these objects.} \\
\end{tabular}
\end{table*}

\setcounter{table}{1}
\begin{table*}
\caption{ Dereddened line intensities of HII regions and PNe, relative to H$\beta$=1.00 \emph{(continued)}}
\label{}\centering\begin{tabular}{lccrrrrrrrrrr}
\hline
ion	 & 	$\lambda$&  f($\lambda$)	 & 	HII\,37	 &	HII\,40	 & 	 PN\,3  	 & 	 PN\,4 	 & 	 PN\,7 	 & 	 PN\,10$^b$  & 	 PN\,11 & 	 PN\,13  &       PN\,14$^c$  & 	 PN\,17   \\
\hline
$[\ion{O}{ii}]$& 	3727	 & 	 +0.255	 & 	 1.55	 &	 1.00	 & 	 0.45: 	 & 	 $<$0.12 & 	 0.94 	 & 	 $<$0.30  & 	  oor	 & 	 $<$0.30  & 	 $<$0.08  & 	 0.34	 \\
H11  	 & 	3770	 & 	 +0.250	 & 	 0.02	 &	 0.02	 & 	 $<$0.08 &	 $<$0.08  & 	 0.04 	 & 	 $<$0.10  & 	 oor 	 & 	 $<$0.20  & 	 0.08 	 & 	 $<$0.03	 \\
H10  	 & 	3798	 & 	 +0.245	 & 	 0.05	 &	 0.03	 & 	 $<$0.08  & 	 $<$0.08  & 	 0.05 	 & 	 $<$0.10  & 	 0.07: 	 & 	 $<$0.20  & 	 0.07 	 & 	 $<$0.030 \\
H9   	 & 	3835	 & 	 +0.236  & 	 0.07	 &	 0.04	 & 	 $<$0.08  & 	 $<$0.08  & 	 0.07 	 & 	 $<$0.10  & 	 $<$0.06 & 	 $<$0.20  & 	 0.08: 	 & 	 0.06: 	 \\
$[\ion{Ne}{iii}]$  & 	3868 & 	 +0.223  & 	0.18	 &	 0.15	 & 	 0.38 	 & 	 0.50 	 & 	 0.39 	 & 	 0.31 	 & 	 0.53 	 & 	 0.51 	 & 	 0.10 	 & 	 0.07	 \\
H8  	 & 	3889	 & 	 +0.223  & 	 0.11	 &	 0.11	 & 	 0.14 	 & 	 0.12 	 & 	 0.15 	 & 	 0.11: 	 & 	 0.11 	 & 	 $<$0.20  & 	 0.12 	 & 	 0.12 	 \\
H7+[\ion{Ne}{iii}] & 	3970 & 	 +0.204  & 	 0.16	 &	 0.16	 & 	 0.31 	 & 	 0.20 	 & 	 0.30 	 & 	 0.14 	 & 	 0.31 	 & 	 0.38: 	 & 	 0.18 	 & 	 0.16	 \\
\ion{He}{i}  & 	4026	 & 	 +0.197  & 	 0.01	 &	 0.01	 & 	 $<$0.03  & 	 $<$0.05  & 	 $<$0.01 & 	 $<$0.08 & 	 0.05 	 & 	 $<$0.20  & 	 0.03 	 & 	 $<$0.03  \\
$[\ion{S}{ii}]$   & 	4069 & 	 +0.187  & 	 0.01	 &	 0.01	 & 	 $<$0.03  & 	 $<$0.05  & 	 0.02 	 & 	 $<$0.08  & 	 $<$0.06 & 	 $<$0.20 & 	 0.02: 	 & 	 $<$0.03 \\
H$\delta$   & 	4102	 & 	 +0.182  & 	 0.16	 &	 0.23 	 & 	 0.23 	 & 	 0.26 	 & 	 0.26 	 & 	 0.22 	 & 	 0.26 	 & 	 0.27 	 & 	 0.26 	 & 	 0.22	 \\
H$\gamma$ & 	4340	 & 	 +0.125  & 	 0.47	 &	 0.47 	 & 	 0.48 	 & 	 0.47 	 & 	 0.47 	 & 	 0.48 	 & 	 0.47 	 & 	 0.45 	 & 	 0.46 	 & 	 0.46	 \\
$[\ion{O}{iii}]$& 	4363 & 	 +0.121  & 	 0.06	 &	 0.05 	 & 	 0.06 	 & 	 0.06 	 & 	 0.12 	 & 	 0.06: 	 & 	 0.14 	 & 	 0.34 	 & 	 0.09 	 & 	 0.05	 \\
\ion{He}{i}& 	4471	 & 	 +0.078  & 	 0.02	 &	 0.03 	 & 	 0.06 	 & 	 $<$0.05 & 	 0.04 	 & 	 0.03: 	 & 	 0.07 	 & 	 $<$0.20 & 	 0.05 	 & 	 0.08	 \\
\ion{He}{ii} & 	4686	 & 	 +0.070  & 	 	 &	  	 & 	 $<$0.03 & 	 0.85 	 & 	 $<$0.01 & 	 $<$0.06 & 	 $<$0.03 & 	 $<$0.20 & 	 0.03 	 & 	 $<$0.03 \\
$[\ion{Ar}{iv}]$\,+\,\ion{He}{i} & 4712	 & +0.039 	 & 	 0.01	 &	 $<$0.01 & 	 $<$0.03  & 	 $<$0.04 & 	 0.02 	 & 	 $<$0.06 & 	 $<$0.03 & 	 $<$0.20 & 	 0.01 	 & 	 $<$0.03 \\
$[\ion{Ar}{iv}]$& 	4738	 & 	 +0.031 	 & 	 $<$0.05 &	 $<$0.01  & 	 $<$0.03  & 	 $<$0.04  & 	 0.01 	 & 	 $<$0.06  & 	 $<$0.03  & 	 $<$0.20 & 	 $<$0.02 & 	 $<$0.03 \\
H$\beta$	 & 	4861	 & 	 +0.000 	 & 	 1.00	 &	 1.00 	 & 	 1.00 	 & 	 1.00 	 & 	 1.00 	 & 	 1.00 	 & 	 1.00 	 & 	 1.00 	 & 	 1.00 	 & 	 1.00	 \\
\ion{He}{i}	 & 	4922	 & 	 $-$0.014 	 & 	 0.01	 &	 0.01 	 & 	 0.04 	 & 	 $<$0.04 & 	 0.02 	 & 	 $<$0.03 & 	 $<$0.02 & 	 $<$0.20 & 	 0.02 	 & 	 $<$0.03 \\
$[\ion{O}{iii}$]& 	4959	 & 	 $-$0.023	 & 	 1.34	 &	 0.99 	 & 	 1.89 	 & 	 1.84 	 & 	 1.83 	 & 	 2.85 	 & 	 2.96 	 & 	 4.80 	 & 	 0.11 	 & 	 1.09 	 \\
$[\ion{O}{iii}]$ & 	5007	 & 	 $-$0.033	 & 	 3.95	 &	 2.87 	 & 	 5.61 	 & 	 5.72 	 & 	 5.59 	 & 	 7.18 	 & 	 8.94 	 & 	 14.20 	 & 	 0.37 	 & 	 4.04 	 \\
\ion{He}{i}	 & 	5017	 & 	 $-$0.036 	 & 	 0.02	 &	 0.01 	 & 	 $<$0.03 & 	 $<$0.04 & 	 0.02 	 & 	 0.03 	 & 	 0.02: 	 & 	 $<$0.20 & 	 0.03: 	 & 	 0.03:	 \\
$[\ion{N}{i}]$ & 	5199	 & 	 $-$0.083 	 & 	 $<$0.01 &	 0.01 	 & 	 0.05 	 & 	 $<$0.04 & 	 0.01 	 & 	 $<$0.05 & 	 $<$0.02 & 	 $<$0.20 & 	 $<$0.04 & 	 $<$0.03 \\
$[\ion{N}{ii}]$& 	5755	 & 	 $-$0.195	 & 	 $<$0.01 &	 $<$0.01 & 	 $<$0.03 & 	 $<$0.04  & 	$<$0.01  & 	 $<$0.05  & 	 $<$0.02 & 	 $<$0.20 & 	 $<$0.03 & 	 $<$0.03 \\
\ion{He}{i}	 & 	5876	 & 	 $-$0.280	 & 	 0.10	 &	 0.10 	 & 	 0.15 	 & 	 $<$0.04 & 	 0.13 	 & 	 0.10 	 & 	 0.13 	 & 	 0.21 	 & 	 0.09 	 & 	 0.12	 \\
$[\ion{O}{i}]$ & 	6300	 & 	 $-$0.284 	 & 	 0.02	 &	 0.01 	 & 	 $<$0.03 & 	 0.02: 	 & 	 0.02 	 & 	 $<$0.05  & 	 $<$0.02 & 	 $<$0.20 & 	 $<$0.02 & 	 $<$0.03 \\
$[\ion{S}{iii}]$& 	6311	 & 	 $-$0.286 	 & 	 0.02	 &	 0.02 	 & 	 $<$0.03& 	 $<$0.04 & 	 0.02 	 & 	 $<$0.05 & 	 0.02: 	 & 	 $<$0.20 & 	 $<$0.02 & 	 $<$0.03 \\
$[\ion{N}{ii}]$& 	6548	 & 	 $-$0.300 	 & 	 0.02	 &	 0.02 	 & 	 0.03 	 & 	 $<$0.04 & 	 0.03 	 & 	 $<$0.05 & 	 0.04 	 & 	 $<$0.20 & 	 $<$0.02 & 	 $<$0.05 \\
H$\alpha$    & 	6562	 & 	 $-$0.322	 & 	 2.86	 &	 2.86 	 & 	 2.85 	 & 	 2.84 	 & 	 2.86 	 & 	 2.76 	 & 	 2.86 	 & 	 2.86 	 & 	 2.86 	 & 	 2.86	 \\
$[\ion{N}{ii}]$& 	6583	 & 	 $-$0.325 	 & 	 0.06	 &	 0.05 	 & 	0.08 	 & 	 $<$0.06  & 	 0.08 	 & 	 0.05: 	 & 	 0.08 	 & 	 0.31 	 & 	 $<$0.03 & 	 0.13	 \\
\ion{He}{i}	 & 	6678	 & 	 $-$0.339 	 & 	 0.03	 &	 0.03 	 & 	 0.04 	 & 	 $<$0.06 & 	 0.03 	 & 	 $<$0.05  & 	 0.04 	 & 	 $<$0.20  & 	 oor 	 & 	 0.06	 \\
$[\ion{S}{ii}]$ & 	6717	 & 	 $-$0.343 	 & 	 0.11	 &	 0.07	 & 	 0.04 	 & 	 $<$0.05 & 	 0.06 	 & 	 $<$0.05 & 	 $<$0.02 & 	 oor  & 	  	 & 	 $<$0.03 \\
$[\ion{S}{ii}]$& 	6731	 & 	 $-$0.344 	 & 	 0.08	 &	 0.06	 & 	 0.03 	 & 	 $<$0.05 & 	 0.06 	 & 	 $<$0.05 & 	 $<$0.02 & 	   & 	  	 & 	 $<$0.03	 \\
\ion{He}{i}	 & 	7066	 & 	 $-$0.383 	 & 	oor	 &	 oor	 & 	 0.05 	 & 	 oor	 & 	 oor 	 & 	 0.11 	 & 	 0.12 	 & 	   	 & 	  	 & 	 0.07	 \\
$[\ion{Ar}{iii}]$ & 	7135	 & 	 $-$0.390 	 & 		 &	 	 & 	 0.04 	 & 	  	 & 	  	 & 	 0.06 	 & 	 0.04 	 & 	 	 & 	  	 & 	 0.02: 	 \\
$[\ion{O}{ii}]$ & 	7320	 & 	 $-$0.410 	 & 	 	 &	 	 & 	 oor 	 & 	  	 & 	  	 & 	 $<$0.07 & 	 0.04 	 & 	  	 & 	  	 & 	 oor	 \\
$[\ion{O}{ii}]$ & 	7330	 & 	 $-$0.411 	 & 	 	 &	   	 & 	  	 & 	  	 & 	  	 & 	 $<$0.07 & 	 0.05 	 & 	  	 & 	 	 & 		 \\
\hline
 c(H$\beta$)& 		 & 		 & 	0.15	& 0.16	 & 	 0.0 	 & 	 0.0 	 & 	 0.63 	 & 	 0.0 	 & 	 0.4 	 & 	 0.0 	 & 	 0.73 	 & 	 0.1 	 \\
\multicolumn{3}{l}{F(H$\beta$) (10$^{-16}$ erg cm$^{-2}$ s$^{-1}$)}  &	45.0 & 	97.7 & 2.30 	 & 	0.99 	 & 	 14.00 	 & 	 3.73 	 & 	 3.50 	 & 	 0.50 	 & 	 4.20 	 & 	 2.80 	 \\
\multicolumn{3}{l}{L(H$\beta$)\,/\,L$_\odot$}		            &    240 &	520  &	12	 & 	5	 & 	74	 &	20	 &	18	 &	3	&	22	 &	15	 \\	
\hline
\multicolumn{12}{l}{
 $b$ Observed with Magellan telescope and spectrograph MIKE at Las Campanas Observatory.}\\
\multicolumn{12}{l}{$c$ The central star shows WR features at $\lambda\lambda$ 4650.  Also \ion{He}{ii} 4686 is of stellar origin.} \\
\end{tabular} 
\end{table*}
\section{Line intensities and reddening}

\subsection{Line intensity measurements}

Calibrated line fluxes were measured for all the available emission lines. The logarithmic reddening correction at H$\beta$, c(H$\beta$), was determined for each object from its H$\alpha$/H$\beta$ ratio, by assuming case B recombination theory (Osterbrock \& Ferland 2006).   Our adopted values of c(H$\beta$) are given in Table 2.  Line ratios relative to H$\beta$ were dereddened  with this c(H$\beta$) and the Seaton (1979) reddening law, assuming a temperature of 10$^4$~K. In general, the values for H$\gamma$, H$\delta$, and higher members of the Balmer series determined this way, were in agreement with the theoretical values to within 15\%.  For a few  objects  the intensities of the higher Balmer lines deviate by more than this, which can be  attributed to faintness of the lines, uncertainties in the flux calibration procedure (we found this could happen for objects in slitlets nearer the CCD  border), anomalous reddening laws, etc.  Since it is important for abundance determinations  that  diagnostic lines such as [\ion{O}{iii}]\,4363 or [\ion{O}{ii}]\,3727 should have values as  reliable as possible, we adjusted the upper Balmer lines to their theoretical values and  corrected all the lines bluer than H$\beta$ accordingly.  

 De-reddened line intensities, on a scale for which $I(\mathrm H\beta)=1.00$, are presented in Table 2.  When a line was not detected, an upper limit  (2 sigma above the noise) is given.  If the line was out-of-range we marked it as `oor'.  For a target with F(H$\beta) \geq $ 3 10$^{-16}$, uncertainties for the line intensities are estimated to be 30\% when the ratio relative to H$\beta$ is about 0.05, 20\% for line intensities $\sim$0.1 H$\beta$. Line intensities larger than 0.1\, H$\beta$ have smaller uncertainties. Very uncertain lines were marked with a colon.  Obviously, uncertainties are better for the brighter objects and poorer for the fainter ones.

The observed fluxes in H$\beta$, F(H$\beta$), as measured through the extraction window, are given at the bottom of Table 2. F(H$\beta$) is the flux from the entire object when its diameter is smaller than 1.7 arcsec (HII\,31 is the only object larger than 1.7 arcsec).  In Table 2, we also list the total de-reddened H$\beta$ luminosity, L(H$\beta$), in solar units, computed by using our c(H$\beta$) and a distance modulus of 25.571$\pm$0.024  (Soszy\'nski  et al. 2006).

\subsection{Extinction}

The foreground extinction towards NGC\,3109 is very low. Burstein \& Heiles (1984)  find E(B-V)=0.04\,mag (c(H$\beta$)=0.06\,dex).  The extinctions determined for various kinds of objects in  this galaxy are also small: Soszy\'nski et al. (2006) report E(V-B)=0.087$\pm$0.012\,mag (c(H$\beta$)=0.12\,dex) for Cepheids while Evans et al. (2007) find  E(V-B)$\leq$ 0.17\,mag for their blue supergiants. The values we obtain for c(H$\beta$) in most of our objects vary from 0.03 to 0.33\,dex (E(V-B) = 0.02 $-$ 0.22\,mag). The only objects showing higher extinction in our sample are  PN\,7  and PN\,14  with c(H$\beta$)$\sim$0.6-0.7\,dex.  Both objects are discussed in more detail later (Sect. 5.1 and Appendix).  Part of the extinction may be intrinsic to these objects.

\subsection{A first glance at the spectra for our PNe and HII regions}

In Fig. 1, the spectra of all the PNe are shown.  As expected, the PNe with better signal-to-noise are the brightest ones: PN\,3, PN\,7, PN\,11 and PN\,14. Many important lines are detected.  In particular,  [\ion{O}{iii}] 4363 was detected with signal-to-noise better than 3 in all objects, allowing us to compute the electron temperature directly.  As expected, the emission lines from  low ionization species (O$^+$, N$^+$, S$^+$) are much weaker in most of the PN candidates (PN\,4, PN\,10, PN\,11, PN\,13, PN\,17) than in HII regions. In some PNe, these lines are even absent.  

It is interesting that, among our PN candidates, only PN\,4 shows emission in \ion{He}{ii} 4686. For all the other PNe, this line is not detected. Although PN\,4 presents a large  He {\sc ii} 4686 / H$\beta$ ratio, He {\sc i} 5876 and the low excitation lines were not detected.  PN\,4 is thus a very high excitation nebula, possibly density bounded. The presence of \ion{He}{ii} in this nebula indicates that it has a central star with a large effective temperature (T$_{\rm eff}$ larger than 70\,000 K).  Among the brightest PNe in  NGC\,3109, it is the only one showing these characteristics.

The  HII regions are much brighter in H$\beta$ than the PNe (by a factor about 10), as expected due to the larger number of ionizing photons from their central stars (see the discussion in Paper I). The excitation as estimated from  the [\ion{O}{iii}]\,5007\,/\,H$\beta$ ratio is lower than in PNe (the only exception is PN\,14, ionized by a cool WR star) and never larger than 3 or 4.  The low ionization species (O$^+$, N$^+$, S$^+$) are always detected. [\ion{O}{iii}]\,4363 is always detected (except for the faint HII\,31 and HII\,34), allowing a direct $T_{\rm e}$-based abundance determination.

\section {Determination of physical conditions, ionic and total abundances}

\subsection{Plasma diagnostics}

Electron temperatures, $T_{\rm e}$, were derived from the [\ion{O}{iii}] 4363/5007 line ratios, and densities, $N_{\rm e}$,  from the  [\ion{S}{ii}] 6731/6717 line ratios, when available.  If not, a density of 1000 cm$^{-3}$ was adopted for PNe and a density of 100 cm$^{-3}$ for HII regions.  These physical conditions were used to determine the ionic abundances of the elements with available lines. The atomic data used are listed in Stasi\'nska (2005), with the exception of the emissivity of the \ion{He}{i} 5876 line, taken from Porter et al. (2007), and the  collision strenghts for [\ion{O}{ii}] lines, taken from Tayal (2007).   The emissivities of 
collisionally excited lines were computed with a 5-level atom approximation, within the code ABELION, written by G. Stasi\'nska.

We assumed that the electron density is uniform in the nebulae  and equal to the one given by the [\ion{S}{ii}] ratio. We also assumed that the electron temperature is uniform and equal to that given by the [\ion{O} {iii}] ratio. Probably, neither of these assumptions is strictly correct. Regarding the electron temperature, for HII regions one could use an empirical relation between $T({\rm O^{+}})$ and $T({\rm O^{++}})$, such as the one proposed by Izotov et al. (2006). For planetary nebulae, on the other hand, there is no clear relationship between $T({\rm O^{+}})$ and $T({\rm O^{++}})$ (G\'orny et al. 2007, in preparation). Applying the rule of Occam's razor, we chose to compute all the ionic abundances with the temperature derived from the [\ion{O}{iii}] ratio. The bias in the computed abundances and ionic ratios due to the assumption of constant temperature and density is likely negligible compared to that due to uncertainties in line intensities.  Furthermore, most of our objects are high excitation nebulae, with O$^{++}$\,/\,O$^+$ $\geq$ 2.  

\begin{table*}
\caption{Electron densities and  temperatures, and ionic abundance ratios (relative to H$^+$) for the PNe and HII regions in NGC\,3109.}   
\label{}\centering\begin{tabular}{lccrrrrrrrr}
\hline
     &   $N_{\rm e}^a$ &     $T_{\rm e}^a$  &     He$^{+}$  &     He$^{++}$  &     O$^{+}$ &    O$^{++}$ &   N$^{+}$ &      Ne$^{++}$ & S$^+$  & Ar$^{++}$ \\
          &   (cm$^{-3}$) &     (K)  &    &      &     ($10^6$) &     ($10^6$) &     ($10^6$) &      ($10^6$) &      ($10^6$)&      ($10^6$) \\
\hline
 HII\,4  &   100. ( 100., 110.)   & 16489. (14873.,17855.)   &  0.082 &        &        &   20.4 &    0.9 &    2.5 &    0.30 &    0.22 \\
 HII\,7  &   100. ( 100., 140.)   & 15295. (14188.,16452.)   &  0.082 &        &   13.7 &   31.6 &    0.6 &    4.3 &    0.24 &    0.30 \\
HII\,11  &   120. ( 100., 470.)   & 14738. (12717.,16806.)   &  0.090 &        &   22.0 &   28.4 &    0.9 &    3.4 &    0.36 &         \\
HII\,15  &   100. ( 100., 310.)   & 13795. (12054.,16054.)   &  0.104 &        &   18.5 &   47.1 &    1.2 &    6.9 &    0.58 &         \\
HII\,17  &                        & 14040. (11576.,16276.)   &  0.084 &        &   20.5 &   34.5 &    1.1 &    2.9 &         &         \\
HII\,20  &   100. ( 100., 620.)   & 13517. (11784.,15016.)   &  0.108 &        &   20.7 &   55.1 &    1.1 &    5.3 &    0.37 &         \\
HII\,30  &   120. ( 100., 730.)   & 11877. ( 7449.,15073.)   &  0.082 &        &   33.9 &   30.8 &    1.0 &    4.0 &    0.36 &    0.31 \\
HII\,32  &   100. ( 100., 170.)   & 13491. (12557.,14626.)   &  0.078 &        &   20.7 &   28.8 &    1.7 &    4.8 &    0.56 &    0.36 \\
HII\,37  &   100. ( 100., 380.)   & 13689. (12506.,14707.)   &  0.079 &        &   17.9 &   55.2 &    0.5 &    6.4 &    0.22 &         \\
HII\,40  &   160. ( 100., 580.)   & 13682. (12474.,15022.)   &  0.078 &        &   11.7 &   40.1 &    0.5 &    5.3 &    0.16 &         \\
PN\,3    &   140. ( 100., 940.)   & 12013. (10624.,13362.)   &  0.116 &$<$0.003 &    7.5 &  113.4 &    0.9 &   17.0 &    0.10 &    0.21 \\
PN\,4    &                        & 11846. (10570.,12743.)   &$<$  0.029 &  0.071 &$<$2.4 &  119.6 &$<$0.7 &   28.8 &$<$0.20 &         \\
PN\,7    &   520. ( 100.,2000.)   & 15542. (14000.,16840.)   &  0.097 &$<$0.001 &    8.4 &   56.1 &    0.6 &    7.8 &    0.12 &         \\
PN\,10   &                        & 10588. ( 9460.,11532.)   &  0.072 &$<$0.005 &$<$8.0 &  213.4 &    0.8 &   27.1 &$<$0.25 &    0.45 \\
PN\,11   &                        & 13495. (11831.,14958.)   &  0.095 &$<$0.003 &        &  128.8 &    0.7 &   19.6 &$<$0.07 &    0.19 \\
PN\,13   &                        & 16484. (12712.,21318.)   &  0.141 &$<$0.018 &$<$2.5 &  123.4 &    2.0 &   10.4 &         &         \\
PN\,17  &   480. ( 450., 500.)   & 12536. (10454.,13814.)   &  0.085 &$<$0.003 &    5.7 &   71.5 &    1.4 &    3.5 &$<$0.11 &    0.11 \\
\hline
\multicolumn{7}{l}{
$a$ In parentheses are given   5th and 95th percentiles  computed by the Monte-Carlo simulation.} \\
\end{tabular}
\end{table*}

In Table 3, we present $T_{\rm e}$, $N_{\rm e}$, and the available ionic abundances for all the objects. For $T_{\rm e}$ and $N_{\rm e}$, we give in parentheses the   5th and 95th percentiles  computed by the Monte-Carlo simulation (see Sect. 4.2).  

The ionic abundances were derived from the following lines (when observed): \ion{He}{i} 5876, \ion{He}{ii} 4686, [\ion{O}{ii}] 3727, [\ion{O}{iii}] 5007, [\ion{N}{ii}] 6583,  [\ion{Ne}{iii}] 3869, [\ion{S}{ii}] 6717+6731, [\ion{Ar}{iii}] 7135 with respect to H$\beta$.  When a line was not detected, its upper limit  was used to compute an upper limit for the corresponding ionic abundance. 

\subsection{Elemental abundances}

The total abundances were calculated from the ionic abundances using the ionization correction factors ({\it icfs}) given by Kingsburgh \& Barlow (1994).    In all of the objects we observed in NGC\,3109, with the exception of PN\,4,  the oxygen abundance is directly  given by the sum of O$^+$ and O$^{++}$ abundances, since no \ion{He}{ii} line is detected.  Thus,  {\it icf}\,=\,1 was used for the O abundance determination for all the objects but PN\,4.  As emphasized by Kingsburgh \& Barlow (1994), the {\it icf} for  Ar is very uncertain. In the case of helium, no effort was made to correct for the neutral species, so that the abundance for this element in low excitation objects is only a lower limit. The weak [\ion{S}{iii}] 6312 line was observed only in a few cases, and given its great sensitivity to the adopted temperature, we do not use it for S$^{++}$ ionic determination.  Therefore, our sulfur abundances are very uncertain. 

The errors in the abundances were derived by Monte-Carlo simulations.  The line intensities were randomly chosen from a gaussian distribution centered on the observed value with a HWHM equal to the estimated uncertainty.   Note that these Monte-Carlo simulations estimate errors in temperatures, densities, and abundance ratios arising only from uncertainties in the reddening-corrected line  fluxes.  They ignore uncertainties due to reddening, as well as systematic uncertainties inherent to our method of computing abundances.  In particular, errors due to the ionization correction factors are not included.  In Table 4, we list the total nominal abundances, as well as the 5th and 95th percentiles  derived from the Monte-Carlo simulation.

The results of our abundance calculations are presented in graphical form in Figs. 3--7 and are analyzed in the following sections.  In all the plots, generally, objects in which at least one coordinate corresponds to an upper or lower limit have been removed for clarity.  However, there are two exceptions to this rule.  The first is when  the true value is likely very close to the limit, as is for example the case for the oxygen abundance in PNe  when O$^+$ is not observed or detected. The other is the oxygen abundance for HII\, 4, which does not take into account the O$^{+}$ ion, and is thus a lower limit.  However, the values of Ne/O and Ar/O for this object are likely as correct as they are for the remaining objects, since they do not depend on the O$^{+}$ abundance.

\begin{table*}
\caption{Computed abundances for the PNe and HII regions in NGC\,3109$^a$.}
\label{}\centering\begin{tabular}{lllllll}
\hline
            &         He/H           &     12+log O/H      &    12+log N/H       &    12+log Ne/H      &     12+log S/H      &    12+log Ar/H      \\
\hline
HII\,4   &  0.082 (0.069,0.093)   &$>$7.31 (7.22,7.40)   &                     &  6.39 (6.27,6.52)   &                     &  5.61 (5.51,5.71)  \\
HII\,7   &  0.082 (0.068,0.100)   &  7.66 (7.57,7.76)   &  6.29 (6.18,6.38)   &  6.79 (6.67,6.90)   &  6.29 (6.20,6.37)   &  5.75 (5.65,5.85)   \\
HII\,11  &  0.090 (0.072,0.116)   &  7.70 (7.55,7.86)   &  6.31 (6.10,6.47)   &  6.78 (6.54,6.98)   &  6.41 (6.31,6.55)   &                     \\
HII\,15  &  0.104 (0.075,0.135)   &  7.82 (7.63,7.98)   &  6.61 (6.46,6.75)   &  6.98 (6.75,7.21)   &  6.68 (6.56,6.82)   &                     \\
HII\,17  &  0.084 (0.054,0.106)   &  7.74 (7.55,7.97)   &  6.48 (6.29,6.70)   &  6.67 (6.42,6.93)   &                     &                     \\
HII\,20  &  0.108 (0.081,0.132)   &  7.88 (7.73,8.04)   &  6.62 (6.44,6.77)   &  6.86 (6.63,7.08)   &  6.49 (6.36,6.62)   &                     \\
HII\,30  &  0.082 (0.054,0.112)   &  7.81 (7.51,8.38)   &  6.30 (6.02,6.69)   &  6.92 (6.18,7.66)   &  6.40 (6.18,6.78)   &  5.77 (5.56,6.18)   \\
HII\,32  &  0.078 (0.066,0.092)   &  7.69 (7.58,7.79)   &  6.61 (6.50,6.69)   &  6.92 (6.73,7.06)   &  6.62 (6.53,6.69)   &  5.83 (5.71,5.94)   \\
HII\,37  &  0.079 (0.067,0.092)   &  7.86 (7.75,7.97)   &  6.30 (6.17,6.45)   &  6.93 (6.80,7.08)   &  6.28 (6.21,6.38)   &                     \\
HII\,40  &  0.078 (0.065,0.087)   &  7.71 (7.58,7.82)   &  6.33 (6.20,6.44)   &  6.84 (6.70,6.98)   &  6.15 (6.05,6.24)   &                     \\
PN\,3    &  0.118 (0.089,0.150)   &  8.08 (7.93,8.26)   &  7.18 (6.96,7.36)   &  7.26 (7.01,7.48)   &  6.20 (6.04,6.39)   &  5.60 (5.44,5.76)   \\
PN\,4    &  0.101 (0.093,0.110)   &  8.44 (8.32,8.62)   &                     &  7.83 (7.70,7.99)   &                     &                     \\
PN\,7    &  0.098 (0.075,0.114)   &  7.81 (7.70,7.92)   &  6.63 (6.47,6.75)   &  6.95 (6.79,7.09)   &  6.14 (6.03,6.27)   &                     \\
PN\,10   &  0.077 (0.059,0.100)   &  8.35 (8.20,8.50)   &$>$7.32 (7.11,7.50)   &  7.45 (7.22,7.66)   &                  &  5.93 (5.75,6.09)   \\
PN\,11   &  0.098 (0.056,0.139)   &  8.11 (7.97,8.28)   &                     &  7.29 (7.01,7.49)   &                     &  5.56 (5.39,5.69)   \\
PN\,13   &  0.158 (0.085,0.229)   &  8.10 (7.82,8.42)   &$>$8.01 (7.74,8.26)   &  7.03 (6.64,7.39)   &                     &                     \\
PN\,17   &  0.088 (0.068,0.110)   &  7.89 (7.74,8.10)   &  7.27 (7.08,7.51)   &  6.57 (6.40,6.83)   & $<$6.21 (6.09,6.38)   &  5.32 (4.53,5.64)   \\
\hline
\multicolumn{7}{l}{$^a$ In parentheses are given   5th and 95th percentiles  computed by the Monte-Carlo simulation} \\
\end{tabular}
\end{table*}

\subsection{The quality of our abundance determinations}

\begin{figure}
\includegraphics[width=\columnwidth]{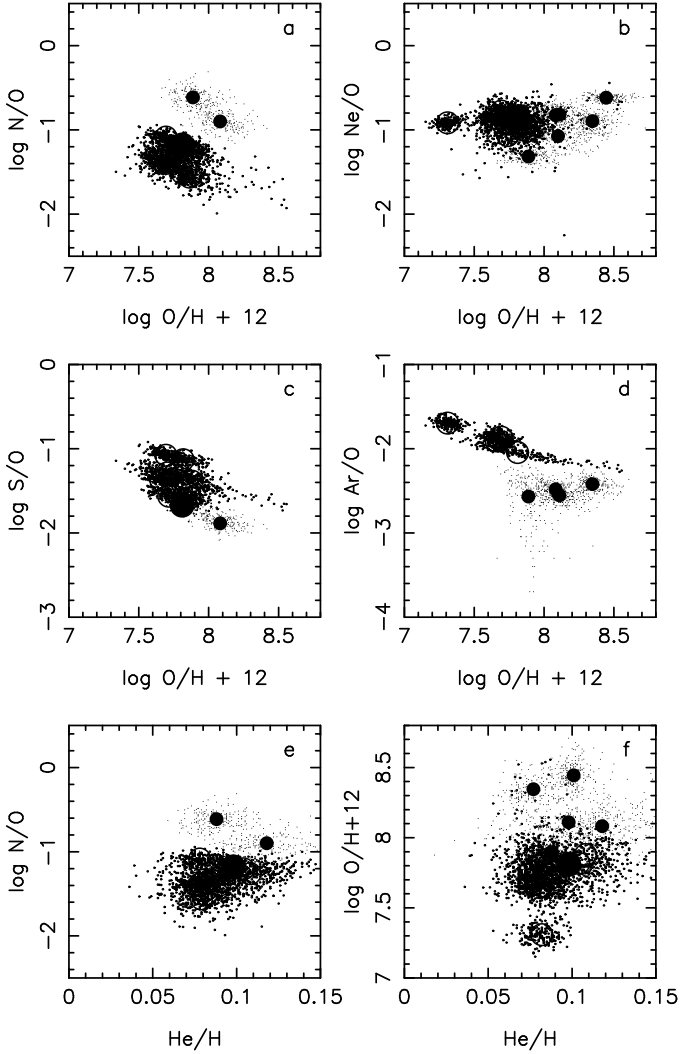} 
\caption{Abundance ratios in PNe (filled dots) and HII regions (open circles) in NGC\,3109. The small points correspond to the results of Monte-Carlo simulations.  \label{fig3}}
\end{figure}

Figure 3 presents various abundance ratio diagrams for the PNe and HII regions in NGC\,3109.  Our Monte-Carlo simulations show that, while the computed Ne/O ratio shows no strong bias with respect to O/H (see Fig. 3b), the N/O, S/O and Ar/O ratios do show a trend (Figs. 3a,c,d): lower N/O, S/O and Ar/O ratios correspond to larger O/H values. This is essentially a consequence of the different temperature sensitivities of the emissivities of the lines used to determine the ionic abundances.  If ignored, these effects may distort the interpretation of  abundance ratio diagrams.  The plots in Fig. 3 illustrate clearly that errors in abundances relative to hydrogen are larger than errors in the abundance ratios of different metals.  Finally, the large errors in the He/H ratios, as compared to the range of variation of this quantity (Figs. 3e,f), preclude any use of the helium abundances for these objects.

\subsection{Abundance determinations for a comparison sample}

We found it useful to compare the results we obtained for PNe and HII
regions in NGC\,3109 with those of corresponding objects in the SMC, a galaxy of similar chemical composition and morphological type. For the PNe, we used the line intensities compiled by Leisy \& Dennefeld (2006), after correction of some obvious errors\footnote{For SMC\,19, we adopted  a null intensity for
[\ion{O}{iii}] 4363, and for SMC\,17  we took  [\ion{S}{ii}]
6731\,/\,H$\beta$ = 1, which makes our abundances more compatible with
the values given by Leisy \& Dennefeld (2006).}, and we recomputed the
abundances in the same way as above. For the HII regions in the SMC, the
situation is far from satisfactory. There is no homogeneous observational data set. We have compiled line intensities from the following references: Peimbert et al. (2000), Kurt et al. (1999), Russell \& Dopita (1990), Heydari-Malayeri et al. (1999), Dufour \& Harlow (1977), retaining  only those HII regions with measurements of [\ion{O}{iii}]\,4363.  This is not expected to lead to a significant bias, since the SMC is metal-poor and the non-detection of this line is likely due to a noisy spectrum rather than to a higher than average metallicity. As for the PNe, we have recomputed the abundances ourselves, after correcting for reddening when necessary. We thus have a sample of 42 PNe and 12 HII regions in the SMC  to compare with our sample of 8 PN candidates and 11 HII regions in NGC\,3109. The average values and the standard deviations of the logarithms of abundance ratios are listed in Table 5 for each group of objects. Note that the standard deviations are computed using the nominal values of the abundance ratio for each object. In this table we also include values for low-metallicity galaxies from Izotov et al. (2006), and the solar values, which we will discuss later.

Figures 4$-$7 present results that will be discussed in the remainder of the paper. All of them follow the same conventions.  Objects in NGC\,3109 are represented by black symbols, objects in the SMC by gray ones. PNe are represented by dots, HII regions by open circles. The PN candidate PN\,7 is represented by a large filled circle, since we have good reason to believe that it is actually an HII region (see below). 

\begin{figure}
\includegraphics[width=\columnwidth]{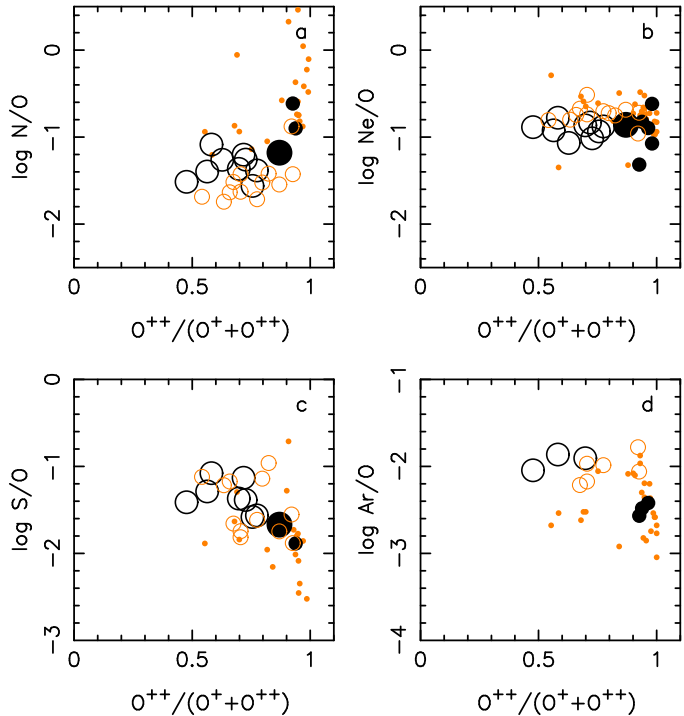} 
\caption{Computed abundance ratios as a function of the nebular excitation. Black: NGC\,3109 objects;  gray: SMC objects. Small filled circles: PNe. Large open circles: HII regions. The large filled black circle is the object  PN\,7 in NGC\,3109. \label{fig4}  }
\end{figure}

\subsection{Abundance ratios versus excitation}

Figure 4 plots various abundance ratios versus nebular excitation, as measured by the O$^{++}$/(O$^{+}$+ O$^{++}$) ratio.  In these diagrams,  PNe show a larger scatter in abundance ratios than HII regions, which can be partially attributed to the fact that the PNe have larger uncertainties because they are fainter, but we consider that part of the scatter is real. The N/O ratio shows much greater scatter among PNe. This is expected since the progenitors of planetary nebulae commonly produce nitrogen. Ne/O is the abundance ratio showing the smallest scatter, both for PNe and HII regions. This ratio also does not show any trend with excitation, which makes us rather confident in the {\it icf} used for O and Ne.  However, the Ne/O ratio in the SMC appears systematically higher (see below).  On the other hand, our values of S/O show a clear trend with excitation, which makes us suspicious of the reliability of the {\it icfs}.  The same may be said for Ar/O, although to a lesser extent. Finally, N/O appears to increase with O$^{++}$/(O$^{+}$+ O$^{++}$), especially for the PNe in the SMC. Most of these objects are of very high excitation, and the relation N/O= N$^{+}$/O$^{+}$ used to derive the N/O ratio is likely systematically biased.  

\section {PNe and  HII regions in NGC\,3109}

\subsection {Reclassification of PN\,7}

\begin{table*}
\caption{Statistics on abundance ratios for the PNe and HII regions in NGC\,3109 and the SMC$^a$.}
\label{}\centering\begin{tabular}{lrrrrrr}
\hline
             &     NGC\,3109 HII$^b$    &      NGC\,3109 PNe$^c$   &       SMC HII        &        SMC PNe       &    HII galx.$^d$ & Sun$^e$ \\
\hline             
log O/H + 12 &    7.77$\pm$0.07 (10)$^f$ &    8.16$\pm$0.19 ( 6) &    8.05$\pm$0.09 (12) &    7.92$\pm$0.32 (30) &  7.77 &  8.66       \\
log N/O      &   $-1.33\pm$0.15 (10) &   $-0.76\pm$0.14 ( 2) &   $-1.51\pm$0.23 (12) &   $-0.54\pm$0.53 (25)&       $-1.42\pm$0.16 &$-$0.88      \\
log Ne/O     &   $-0.90\pm$0.08 (11) &   $-0.98\pm$0.19 ( 6) &  $-0.73\pm$0.10 (12) &  $-0.75\pm$0.22 (30) & $-0.78\pm$0.07  &  $-$0.82     \\
log S/O      &   $-1.38\pm$0.20 ( 9) &   $-1.90\pm$0.00 ( 1) &   $-1.47\pm$0.32 (12) &   $-1.84\pm$0.45 (17) & $-1.72\pm$0.11  & $-$1.52     \\
log Ar/O     &   $-1.94\pm$0.14 ( 3) &   $-2.50\pm$0.10 ( 4) &   $-2.03\pm$0.16 ( 6) &   $-2.46\pm$0.31 (28) &    $-2.38\pm$0.10 & $-$2.48     \\
log He/H     &   $-1.07\pm$0.06 (11) &   $-1.06\pm$0.11 ( 6) &   $-1.10\pm$0.09 (12) &   $-1.03\pm$0.12 (30) & &  -1.07      \\
\hline
\end{tabular}
\begin{tabular}{l}
$a$ In parentheses are given the number of objects. \\
$b$ With PN\,7. \\
$c$ PN\,7 excluded. \\
$d$ Values for low metallicity galaxies from Izotov et al. (2006). Ne/O, S/O and Ar/O were derived from their eq. 30, 26 and 31 for O/H=7.77.\\ N/O is an estimate  derived by ourselves.\\
$e$ The solar abundances are the photospheric values taken from Grevesse, Asplund \& Sauval (2007). \\
Note that the solar abundances of He, Ne and Ar are more uncertain than those of the other elements.\\
$f$ HII\,4 excluded. \\

\end{tabular}
\end{table*}

Figure 5a plots the total H$\beta$ luminosity, L(H$\beta$), corrected for extinction as a function of O/H for all the objects in NGC\, 3109 for which we obtained spectra. Clearly, our HII regions are grouped at higher H$\beta$ luminosities,  while the PN candidates are found at lower H$\beta$ luminosities, with a distinct gap between the two classes, from about 20 to 60 L$\odot$. However, one PN candidate, PN\,7, represented by a large filled circle, has a luminosity similar to that of our HII regions and much higher than that of the remaining  PN candidates.  PN\,7's H$\beta$ luminosity thus provides a reason to question its status as a PN. In the light of [\ion{O}{iii}] 5007, PN\,7 is also 1\,mag brighter than the next brightest PN (Paper I) and surprisingly bright compared to the luminosity expected for its metallicity (Ciardullo et al. 2002).  Also its central star is much brighter than expected for PNe. 
On the other hand, its density and excitation, as measured via the [\ion{O}{iii}] 5007 intensity, are  higher than for the other H II regions in NGC\,3109 and more similar to the values found in the PNe (note that the strength of its [\ion{S}{ii}] 6716,6731 lines make it doubtful that this object is matter-bounded). However, such densities and excitations are  not unusual in HII regions in other galaxies,  so these cannot be taken as arguments for it being a PN. Although spectroscopy has  not clearly resolved the ambiguity remaining from the photometry of PN\,7,  we  consider PN7 an HII region on the basis of its stellar and nebular luminosity being definitely much higher than the other PNe.

Our reclassification of  PN\,7 as an HII region implies that the true cumulative PN luminosity function in NGC\,3109 has a  peak absolute magnitude of $-$3.93 mag, a value that is slightly faint compared to the value of $-$4.08 mag predicted by Ciardullo et al. (2002) for galaxies with metallicities of about  log~O/H\,+\,12 $\sim$ 8.0.  However, the small number of PNe in NGC\,3109 make it unlikely that a PN would be observed at the maximum luminosity.  If our interpretation of PN\,7 as an HII region is correct, the peak absolute magnitude of the planetary nebula luminosity function (PNLF) continues to decrease at lower metallicities. 

\begin{figure}
\includegraphics[width=\columnwidth]{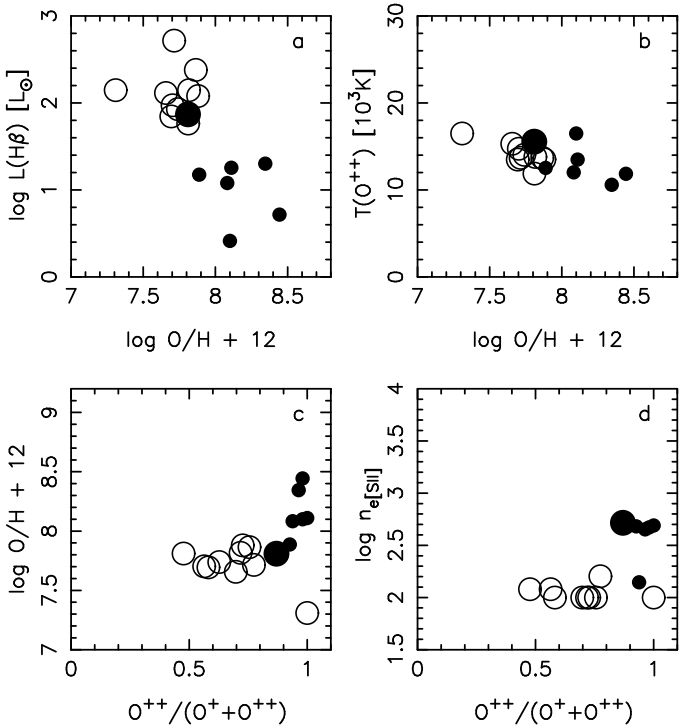} 
\caption{Comparison of characteristics in PNe (small filled circles) and HII regions (open circles) in NGC\,3109. The large filled circle is the object  PN\,7. The large circle appearing at the lowest O/H is HII\,4, for which we can derive only a lower limit to the oxygen abundance, since the [\ion{O}{ii}]\,3727 line was out of the observed wavelength range.  \label{fig5} }
\end{figure}

\subsection{Differences between the chemical composition of PNe and HII regions in NGC\,3109}

\begin{figure}
\includegraphics[width=\columnwidth]{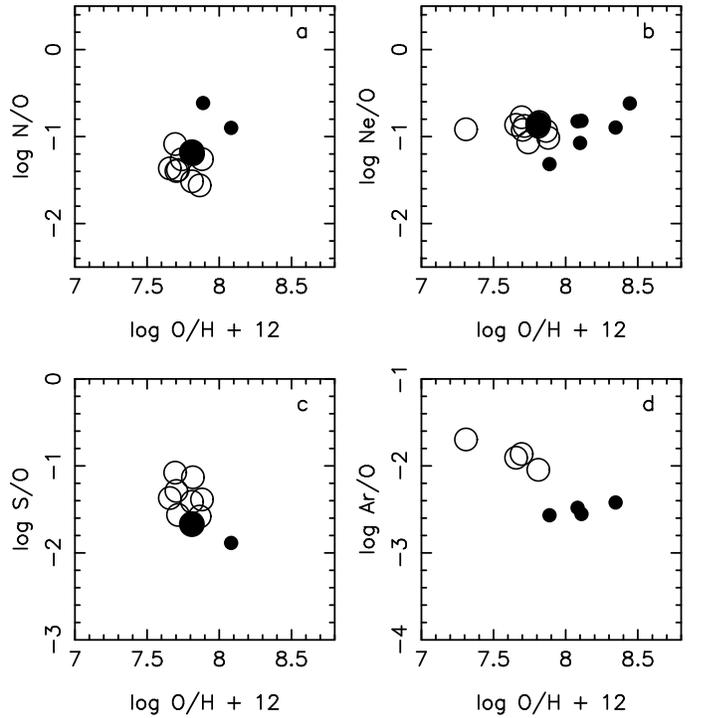} 
\caption{Comparison of abundance ratios between PNe and HII regions in NGC\,3109. The large circle appearing at the lowest O/H is HII\,4 ( see caption in Fig. 5).  \label{fig6}}
\end{figure}

In Fig. 6, we compare the abundance ratios between HII regions and PNe in NGC\,3109. The HII regions show a very small spread in their oxygen abundances with an average value  log~O/H\,+\,12\,=\,7.77$\pm$0.07\,dex  (including PN\,7). This is in remarkable agreement with the value  7.76$\pm$0.07\,dex recently obtained from an analysis of 8 B supergiants by  Evans et al. (2007) in a region covering 4\arcmin. The small scatter in our oxygen abundances, throughout a much larger region,  indicates that the present ISM in NGC\,3109 is chemically very homogeneous.  Previous studies of HII regions in this galaxy (Lee et al. 2003;  Kniazev et al. 2006), have found similar oxygen abundances, but were unable to verify the homogeneity because they only had one or two objects with reliable abundances.  

All of the PNe for which we could determine the oxygen abundance have an O/H ratio larger than that for any of  the HII regions, as is clearly seen  in Figs. 5a-c and 6b\footnote {Note that in all our PNe (but  PN\,4), the values of O/H were calculated directly adding O$^+$ and O$^{++}$, so that the larger values found for O/H cannot be ascribed to an incorrect  {\it icf}.}. The difference in the average values of log O/H is 0.39 dex. A similar situation occurs in Sextans A, where the HII regions have a mean oxygen abundance of 12 + log (O/H) = 7.6$\pm$0.2\,dex, while the only known PN has an oxygen abundance 0.4 dex higher (Kniazev et al. 2005; Magrini et al. 2005). The latter authors (and also Leisy et al. 2002; Leisy \& Dennefeld 2006) suggested that PNe with higher oxygen abundance than the HII regions in the same environment might have experienced  oxygen dredge-up.  This is predicted by theoretical models for the evolution of low metallicity (Z $\leq$ 0.25Z$_\odot$) PN progenitors (Marigo 2001; Herwig 2004), especially those in the 2-3 M$_\odot$ range. Our study of PNe and HII regions in NGC\,3109 supports their suggestion, and strenghtens it considerably.

In the HII regions of NGC\,3109, the Ne/O ratio also shows very little dispersion. We find log Ne/O = $-0.90\pm$0.08\,dex,  lower  by 0.12\,dex than the value found by Izotov et al. (2006) for their  sample of low metallicity galaxies (see Table 5).  Despite their uncertainties, the  Ar/O ratios are also remarkably uniform in the  HII regions of NGC\,3109.   The  values of both Ar/O and S/O  are higher than predicted from the Izotov et al. (2006) relation, by more than twice the dispersion about this relation in the case of Ar/O.   It should be noticed however, that Izotov et al.'s calculations were done with an {\it icf} for Ar very different from the one we have employed here. If we had used the {\it icf} of  Izotov et al. our Ar abundances  would be lower by a factor  of almost two, and our Ar/O average, closer to Izotov et al.'s value. 

In the PNe in NGC\,3109, the  average Ne/O ratio is equal, within uncertainties, to that in the HII regions, although with a  larger dispersion ($\pm$ 0.19 dex), as seen in Fig. 6b and Table 5.  As Fig. 3 shows, this larger dispersion is partially a consequence of observational uncertainties.   Since we have argued that fresh oxygen must have been dredged up in the PNe in NGC\,3109, this would be also true for neon, at least for some objects.   It is puzzling though, that this neon and oxygen production would have maintained the abundance ratio produced by supernovae in the ISM.

There are few PNe in NGC\,3109 for which the S/O and Ar/O  ratios could be calculated  (Figs. 6c,d and Table 5).  The Ar/O average for the PNe is significantly  lower than the average in the HII regions. This appears consistent with the idea that the PNe dredged up oxygen, by more than a factor of 0.3\,dex, but the large uncertainties in the Ar abundances determination (mainly due to the uncertainties in the {\it icfs})  complicates the interpretation of this result.   

The  value of log N/O  also shows small dispersion  in the HII regions ($-1.31\pm$0.15) and is significantly lower than the solar value of Grevesse et al. (2007). However, similar values have been reported for low  metallicity HII galaxies (Izotov et al. 2006).  For the two confirmed PNe in NGC\,3109, we obtain values of N/O larger than in HII regions, but this is not unexpected since the progenitors of PNe commonly synthesize and dredge-up nitrogen.

\section {NGC\,3109 versus the SMC}

\subsection {Abundance patterns in HII regions}

\begin{figure}
\includegraphics[width=\columnwidth]{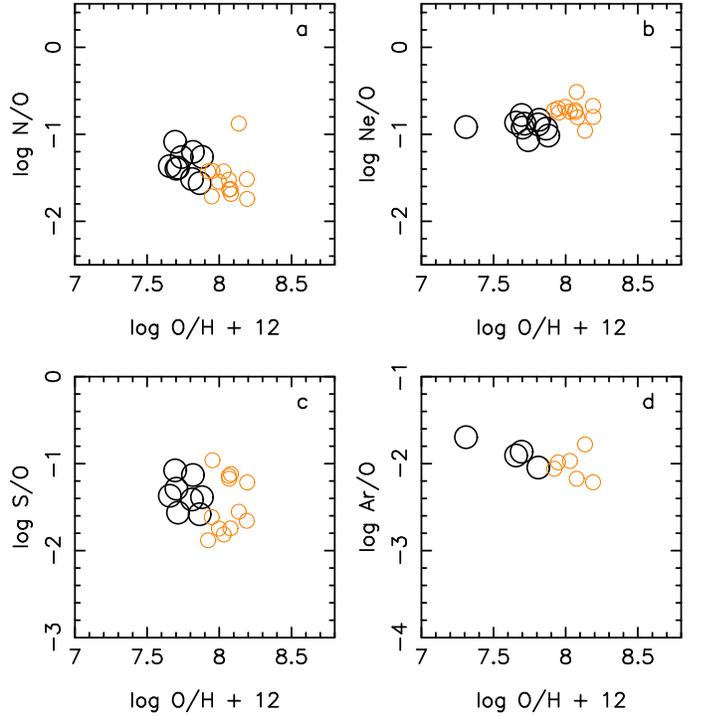} 
\caption{Comparison of abundance ratios  in the HII regions of NGC\,3109 (black symbols) and the SMC (gray  symbols). The large circle appearing at the lowest O/H is HII\,4, for which we can derive only a lower limit to the oxygen abundance, since the [\ion{O}{ii}]\,3727 line was out of the observed wavelength range.  \label{fig3}}
\end{figure}

Figure 7 shows the N/O, Ne/O, S/O and Ar/O abundance ratios as a function of log~O/H\,+\,12 for the HII regions in NGC\,3109 and in the SMC.
In both galaxies, there is very little dispersion among the oxygen abundances in the HII regions. The ISM in NGC\,3109 is clearly less enriched than that in the SMC. 

The ISM in both galaxies is also homogeneous in N/O. There is marginal evidence from Table 5 that the N/O ratio is higher in the HII regions in NGC\,3109 than in  those in the SMC, by  0.17 dex on the average. Figure 7a suggests that the difference is real. Although it is somewhat curious that the SMC has the lower N/O ratio, a comparison with the values typically found in dwarf irregular galaxies (Izotov et al. 2006) indicates that the values found in both galaxies are not unusual for their metallicities.  Presumably, the difference in N/O is the result of different histories of star formation in the two galaxies, which is not surprising.

The Ne/O ratios also show very little dispersion in the HII regions in both NGC\,3109 and the SMC.  The average of Ne/O appears smaller by 0.18 dex in NGC\,3109.  This dependence of Ne/O upon O/H is much larger than  found by Izotov et al. (2006) in blue compact dwarf galaxies and interpreted by them as being due to oxygen depletion onto dust grains in more metal-rich galaxies.  We do not have a convincing explanation for this fact so far.

The values of S/O show a very large scatter.   Since the sulfur abundances were derived from  the [\ion{S}{ii}] 6716+31 lines only, they cannot be considered reliable, and we do not discuss them further.

The Ar/O ratio appears to decrease as O/H increases (see Fig. 7d).   However, the mean values in both galaxies are equal and error propagation could produce this trend, as seen in Fig. 3d.   Note that the appearance of a trend is enhanced by the HII region with the lowest O abundance in NGC\,3109 (HII\,4).  The oxygen abundance for this object, however, is only a lower limit, since [\ion{O}{ii}] 3727 was not observed. Therefore, our data are compatible with a constant value of Ar/O in the ISM in both galaxies, as expected.  

\subsection {Abundance patterns in PNe}

\begin{figure}
\includegraphics[width=\columnwidth]{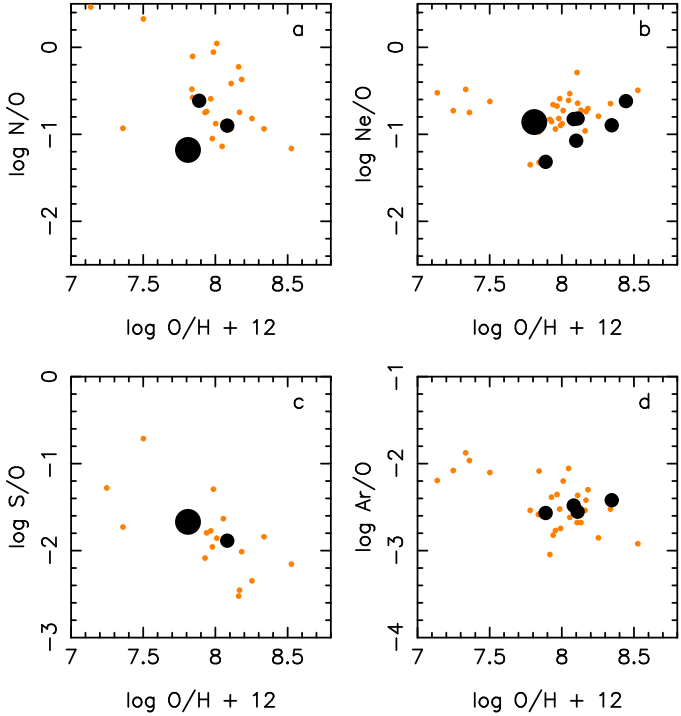} 
\caption{Comparison of abundance ratios between PNe in NGC\,3109 (black symbols) and PNe in the SMC (gray symbols).  \label{fig3}}
\end{figure}

Figure 8 presents the N/O, Ne/O, S/O and Ar/O ratios derived for the PNe in both galaxies and for PN\,7, represented by the large black dot.  In general, a much larger dispersion is seen in this figure than in Fig. 7.  There are few PNe from NGC\,3109 in this diagram.  However,  their N/O, Ne/O, S/O and Ar/O ratios are similar to those of the PNe in the SMC. 

The two PNe in NGC\,3109 for which we could derive the nitrogen abundance are not extremely nitrogen-rich.  It is not excluded that, in a larger sample of PNe with nitrogen abundances, one might find a few nitrogen-rich PNe. Note, however, that the SMC PNe with the largest values of N/O are all of very high excitation (see Fig. 4a) and it is possible that their nitrogen abundances have been overestimated.  On the other hand, Richer \& McCall (2007) found that bright PNe in dwarf irregulars often have N/O ratios similar to those found in the ISM in these galaxies, so our results for NGC\,3109 may not be unusual. 

The range in oxygen abundances for the SMC PNe is much larger than that for the PNe in NGC\,3109 (compare Fig. 8 and Fig. 5a, where all the PNe in NGC\,3109 are plotted). In particular, in NGC\,3109, there are no objects with log O/H +12 $<$ 7.8\,dex, as in the case of the SMC.  This could merely be an effect of poor statistics.  However, it is also possible that the population of PNe in NGC\,3109 is truly devoid of objects that experienced ON cycling, transforming oxygen into nitrogen, or is entirely composed of objects enriched in oxygen.

\subsection{ The evolution of PNe in NGC\,3109 and in the SMC}

\begin{figure}
\includegraphics[width=\columnwidth]{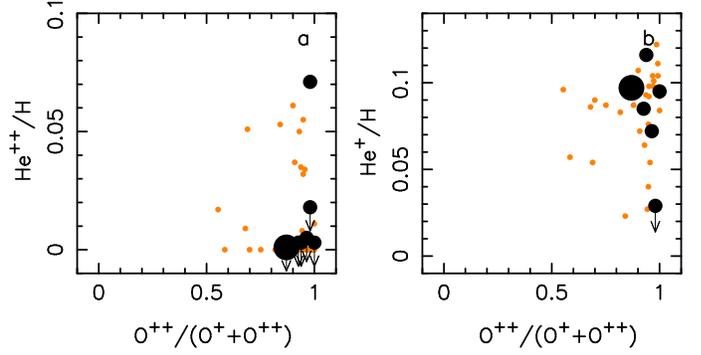} 
\caption{The ionization of helium as a function of   O$^{++}$/(O$^{+}$+ O$^{++}$) for PNe in NGC\,3109 and in the SMC.  \label{fig3} }
\end{figure}

Figures 9a,b display the values of  He$^{++}$/(He$^{+}$+ He$^{++}$)  and He$^{+}$/(He$^{+}$+ He$^{++}$), respectively,  versus  O$^{++}$/(O$^{+}$+ O$^{++}$). It is remarkable that, although all our PNe in NGC\,3109 are of high excitation (as measured by  O$^{++}$/(O$^{+}$+ O$^{++}$), in only one is the \ion{He}{ii} 4686 line detected.  This is not a matter of bad signal-to-noise since lines  with intensities below 0.05 H$\beta$ have been detected, so these PNe have no detectable \ion{He}{ii} 4686 line.   The very different distribution of line intensities in the PNe in NGC\,3109 and in the SMC argues for a significantly different population of bright PNe in the two galaxies. As seen in Fig. 3 of Stasi\'nska et al. (1998), differences in central star masses, nebular masses or expansion velocities can induce different distributions in the  He$^{++}$/(He$^{+}$+ He$^{++}$)   vs.   O$^{++}$/(O$^{+}$+ O$^{++}$) diagram.  

Stasi\'nska et al. (1998) found that PNe in different galaxies had different distributions in such diagrams. For example, most of the PNe in both Magellanic Clouds have \ion{He}{ii}\ $4686/\mathrm H\beta$ larger than 0.2, while, in more metal-rich environments, such as the Galactic bulge or the bulge of M31, a large fraction of the PNe present \ion{He}{ii}\ $4686/\mathrm H\beta$ lower than that.  This seemed to indicate a trend between the distribution of the \ion{He}{ii}\ $4686/\mathrm H\beta$ ratio and the metallicity of the parent galaxy.  NGC\,3109, which is more metal-poor than those galaxies, does not follow such a trend.  Another possibility is that the ages (or masses) of the stellar progenitors of the PN population is responsible for establishing the distribution of \ion{He}{ii}\ $4686/\mathrm H\beta$ ratios.  In either case, the effect does not appear to be linear.  Perhaps this result is biased by small number statistics and further studies of PNe in NGC\,3109 and other metal-poor galaxies should investigate this point. 

\section{The oxygen abundance discrepancy in NGC\,3109: dredge-up in PN progenitors or infall of metal-poor gas?}

 In Sect. 5.2 we argued that the abundance differences between the PNe and the HII regions in NGC\,3109 seemed to indicate that the PNe in NGC\,3109 arise from stellar progenitors that have synthesized and dredged-up oxygen. However, an alternative explanation for the apparent enrichment of oxygen and neon in the PNe in NGC\,3109 is that the ISM in NGC\,3109 could have  been diluted in the recent past.  Based upon its kinematics, Barnes \& de Blok (2001) suggest that NGC\,3109 could have been perturbed by an interaction with its neighboring galaxy, Antlia, about one Gyr ago.  Based upon their \ion{H}{i} measurements, the ratio of \ion{H}{i} mass to blue luminosity, $M_{H\ I}/L_B$, for NGC\,3109 is somewhat high compared to the typical values observed in dwarf irregular galaxies, 2.5 rather than about 1.0 in solar units (Roberts \& Haynes 1994).  Given the metallicity found here for its ISM and the distance we adopt, NGC\,3109 is over-luminous for its oxygen abundance in a metallicity (oxygen abundance) versus luminosity diagram, regardless of whether blue or mid-infrared luminosities are considered (Lee et al. 2003; Lee et al. 2006).  The oxygen abundance deficit is about 0.2\,dex.  The magnitudes of the over-abundance of \ion{H}{i} and the oxygen deficit are in reasonable agreement with the difference in abundance between the PNe and HII regions in NGC\,3109.  It seems feasible, then, to explain the three results simultaneously if the present ISM in NGC\,3109 were diluted in the recent past, via the injection of very low metallicity gas or if lower metallicity gas from its outer regions had been mixed with that in its inner regions, perhaps during its putative encounter with its companion, the Antlia galaxy.  There is, however, an important  problem with this scenario.  An enormous amount of gas is required (at least half of the present-day gas mass) to dilute the metallicity of  NGC\,3109  by a factor of two.  It seems unlikely that the gas mass needed for such extreme dilution would have been available.   

\section{Main conclusions}

We have presented spectroscopic data for 20  nebulae  in NGC\,3109 that we had previously classified as candidate PNe or compact HII regions using photometric data (Paper I).   From the calibrated spectra we have derived the physical conditions, $T_{\rm e}$ and $N_{\rm e}$,  and computed the abundances of He, O, Ne, N, S and Ar, using the classical empirical method based upon the determination of T$_{\rm e}(\ion{O}{iii})$. 

Our spectroscopic study confirms our previous classification into PNe and HII regions, with the exception of one object, PN\,7, which we had already indicated as problematic. For this object,  the H$\beta$ luminosity, the stellar magnitude and  the location in all the abundance ratio diagrams argue for an identification as an HII region.  With only one  uncertain classification for 20 objects, the classification procedure employed in Paper I, combining photometric ratios and luminosities, is a remarkable success.  On the other hand, because such a procedure has not been systematically used previously, we note that a number of objects considered to be extragalactic PNe might actually be compact HII regions masquerading as PNe. This obviously has implications on the use of PNe as standard candles for determinations of galaxy distances (see Ciardullo 2005 for a recent review) or as tracers of the stellar background in galaxies or in the intergalactic medium (Buzzoni et al. 2006). 

  We find that the  HII regions in NGC\,3109, have remarkably uniform abundances across the galaxy. The dispersion in the  abundance ratios is of the order of $\pm 0.15$\,dex or less for all the elements considered. Given that our objects span the entire face of the galaxy, this implies that the  ISM in NGC\,3109 is very homogeneous.  Our oxygen abundances  are very similar to those obtained by former studies for a smaller number of  HII regions (Lee et al. 2003;  Kniazev et al. 2006) and to the abundances derived for 8 B supergiants by  Evans et al. (2007). 
 This is the first time that the large scale chemical homogeneity of the ISM is demonstrated for such a metal-poor galaxy.  For the purposes of comparison, we derived the chemical composition of HII regions in the SMC  and find that NGC\,3109 is clearly more metal-deficient.

In both NGC\,3109 and the SMC, the N/O and Ne/O ratios in HII regions show very little scatter.  In the SMC, both  are in agreement  with what is typically found for metal-poor star-forming galaxies (Izotov et al. 2006).   In NGC\,3109, N/O is in agreement with the results of Izotov et al. (2006), but Ne/O  is lower than expected by 0.12 dex.  We have found no convincing explanation  for this deficit.

The PNe, on the other hand, show a significantly different abundance pattern from the HII regions in NGC\,3109. Their oxygen abundances are larger, by 0.39 dex on average, than those found in the HII regions. This favors models  predicting oxygen dredge-up in low-metallicity PN progenitors.  The Ne/O  ratios are more scattered than in HII regions (partially as a result of observational uncertainty), but have the same average value (within incertainties) as the H II regions. This would imply that neon was also affected by the evolution of the PN progenitors. 

On average, PNe have  lower oxygen abundances than HII regions in the SMC, by 0.13 dex. In NGC\,3109, the difference is larger and in the opposite direction. This indicates that, while evidence for oxygen dredge-up was seen only for the most metal-poor PNe in the SMC (Leisy \& Dennefeld 2006), it  seems clear for all the PNe in NGC\,3109. As the neon abundance in the PNe of  NGC\,3109 is apparently also affected by stellar evolution,  the above would imply that neither oxygen nor neon in PNe  are always  reliable indicators of the chemical composition of the ISM at  the low metallicity of NGC\,3109. 

An alternative to the  conclusion that PNe in NGC\,3109 have synthesized and dredged up oxygen and neon is that its ISM has been diluted in the recent past with low metallicity gas.  While a number of facts support  this conjecture, it seems unlikely that the amount of low-metallicity gas required to dilute the ISM,  at least half the present-day gas mass, would have been available.

Therefore,  it seems so far that the best explanation for the difference in chemical composition in PNe and HII regions in NGC\,3109 is that oxygen and neon have been synthesized and dredged-up during the evolution of the progenitors of these low metallicity PNe.   To confirm this hypothesis, it would be fruitful to observe a larger sample of PNe over a larger wavelength range in order to derive more reliable S and Ar abundances.  In addition, the {\it icfs}  for these crucial elements need to be improved.

The excitation patterns of the PNe in NGC\,3109 and the SMC are very different. A similar situation was already noted by Stasi\'nska et al. (1998) in a comparative study of PNe in five galaxies of different metallicities.  From that study it seemed that more metal-poor galaxies had a larger proportion of PNe with high  \ion{He}{ii} 4686\,/\,H$\beta$ ratios,  whereas NGC\,3109 does not follow this trend. This is one more puzzle to consider in unravelling the evolution of PNe.

\begin{acknowledgements}
M. Pe\~na is grateful to DAS, Universidad de Chile, for hospitality during a sabbatical stay when part of this work was performed. M.P. gratefully acknowledges financial support from FONDAP-Chile and DGAPA-UNAM.  This work received financial support from grants \#43121 (CONACYT-Mexico), IN-108506-2, IN-108406-2, IN-112103 and IN-114805 (DGAPA-UNAM). G.S. benefited from the hospitality of the Instituto de Astronomia (UNAM) and from financial support from DGAPA-UNAM and CONACYT-Mexico.
\end{acknowledgements}

{\bf Appendix}

1. PN\,4  is the only PN where \ion{He}{ii} 4686 is detected.  It has a large \ion{He}{ii} 4686\,/\, H$\beta$ ratio while \ion{He}{i} 5876 is not detected. Therefore, it is a very high excitation PN. In paper I, this object was one of the PN candidates  undetectable in the {\it r}-band image, indicating a large [O {\sc iii}]/H$\alpha$ ratio. The {\it icf} for O in this nebula should be large due to its high degree of ionization (a lot of O should be in the O$^{+3}$ and O$^{+4}$ stages). The oxygen abundance given  in Table 4 is a very uncertain, since no reliable {\it icf} could be computed because \ion{He}{i} 5876 intensity is only an upper limit. 

2. PN\,14 has a WR central star.  Its spectrum shows prominent and wide features at $\lambda$4645 and  $\lambda$4686.  The FWHM of these lines is 30 \AA, the only features in any of the spectra exceeding the instrumental resolution of about 8 \AA.  No other stellar lines are clearly detected due to the stellar faintness. WR features and stellar emission lines have been detected in several extragalactic PN central stars, in particular in the dwarf spheroidal galaxies Sagittarius and Fornax (Zijlstra  et al 2006; Kniazev et al. 2007). The low ionization degree of the nebula ([\ion{O}{iii}] 5007\,/\,H$\beta$ is lower than 1) indicates a low T$_{\rm eff}$ central star. The [OIII]4363\,/\,5007 ratio is completely anomalous for a normal PN, indicating a large electron density  (probably Ne $>>$ 10$^4$ cm$^{-3}$).  Unfortunately the density-sensitive [\ion{S}{ii}] lines  fell beyond the range of our wavelength coverage.   PN\,14 appears similar to some galactic planetaries like  CPD$-56^o$8032 or He\,2-113,  which have very large densities and whose [WC] central stars present effective temperature of about 30,000 K (DeMarco \& Crowther 1998).  This nebula presents a reddening higher than the average  (c(H$\beta$)=0.72), probably indicating the presence of significant quantities of dust in the nebula.

\end{document}